\documentclass{aa}
\usepackage{psfig}
\usepackage{txfonts}
\usepackage{graphicx}
\usepackage{amssymb}
\usepackage[below]{placeins}
\usepackage{natbib}
\bibpunct{(}{)}{;}{a}{}{,}

\begin{document}

\title{XMM-Newton Spectroscopy of the Cluster of Galaxies 2A~0335+096}
\author{N. Werner\inst{1}
 \and J. de Plaa\inst{1,2}
 \and J.S. Kaastra\inst{1}
 \and Jacco Vink\inst{1,2}
 \and J. A. M. Bleeker\inst{1,2}
 \and T. Tamura\inst{3}
 \and J. R. Peterson\inst{4}
 \and F. Verbunt\inst{2}}

\offprints{N. Werner, email {\tt n.werner@sron.nl}}

\institute{     SRON Netherlands Institute for Space Research, Sorbonnelaan 2,
                NL - 3584 CA Utrecht, the Netherlands
         \and
                Astronomical Institute, Utrecht University, P.O. Box 80000,
                NL - 3508 TA Utrecht, the Netherlands
	 \and
	 	Institute of Space and Astronautical Science, JAXA, 3-1-1 Yoshinodai, Sagamihara, Kanagawa 229-8510, Japan
	 \and
		KIPAC, Stanford University, PO Box 90450, MS 29, Stanford, CA 94039, USA}

\date{Received 20 July 2005, Accepted 30 November 2005}

\abstract{We present here the results of a deep (130 ks) XMM-Newton observation of the cluster of galaxies 2A~0335+096. 
The deep exposure allows us to study in detail its temperature structure and its elemental abundances. 
We fit three different thermal models and find that the multi-temperature {\it{wdem}} model fits our data best.
We find that the abundance structure of the cluster is consistent with a scenario where the relative
number of Type~Ia supernovae contributing to the enrichment of the intra-cluster medium is $\sim25$\%, while the relative number of core collapse supernovae is $\sim75$\%.  
Comparison of the observed abundances to the supernova yields does not allow us
to put any constrains on the contribution of Pop~III stars to the enrichment of the ICM.
Radial abundance profiles show a strong central peak of both Type~Ia and core collapse supernova products. Both the temperature and iron abundance maps show an asymmetry in the
direction of the elongated morphology of the surface brightness. In particular the temperature map shows a sharp change over a brightness edge on the southern side of the core,
which was identified as a cold front in the Chandra data. This suggests that the cluster is in the process of a merger with a subcluster. Moreover, we find that the blobs or
filaments discovered in the core of the cluster by Chandra are, contrary to the previous results, colder than the ambient gas and they appear to be in pressure
equilibrium with their environment.  

}

\maketitle

\section{Introduction}
\label{intro}

Clusters of galaxies are the largest known gravitationally bound structures in the universe. 
According to the standard cosmological scenario they form and grow along the filaments through
merging with groups and individual galaxies. Optical and X-ray studies reveal that clusters of galaxies are still forming at the present epoch.

The large effective area and superb spectral resolution of \emph{XMM-Newton} together with the high spatial resolution of \emph{Chandra} allow us to study
clusters of galaxies with unprecedented detail. Analysis of data obtained by these satellites led to a number of important results in the recent years.
In the central parts of many clusters of galaxies the gas density is high enough that the radiative cooling time of the gas is shorter than the age of the cluster.
As the gas cools the pressure decreases, which causes a net inflow toward the center of the cluster. Many clusters of galaxies indeed show a temperature
drop by a factor of three or more within the central 100 kpc radius \citep[for a review on cooling flows see][]{fabian1994}. However, the spectra obtained by the XMM-Newton
Reflection Grating Spectrometer (RGS) show no evidence for strong cooling rates of gas below $30-50$\% of the maximum temperature of the ambient gas, which forces us to look for
additional heating mechanisms in the cores of clusters \citep[]{peterson2001,tamura2001a,kaastra2001}. The high resolution images from Chandra led recently to the discovery of
{\it{cold fronts}}, associated with motion of the cluster cores and to the identification of filamentary structure in the cores of a number of clusters \citep{Markevitch2000}.
Spatially resolved spectroscopy of many clusters of galaxies shows a strongly centrally peaked distribution of metal abundances \citep{tamura2004}.

Because of their large potential wells, clusters of galaxies retain all the
enriched material produced in the member galaxies. This makes them a unique environment for elemental-abundance measurements and for the study of the
chemical enrichment history of the universe. 

In this paper we study the X-ray bright cluster 2A~0335+096 using spatially-resolved and high-resolution spectra obtained during a 130 ks observation 
with the European Photon Imaging Camera \citep[EPIC,][]{turner2001,struder2001} and the Reflection Grating Spectrometer
\citep[RGS,][]{herder2001} aboard \emph{XMM-Newton} \citep{jansen2001}. The properties of 2A~0335+096 allow us to address here some of the above mentioned issues.

2A~0335+096 was first detected as an X-ray source by \emph{Ariel V} \citep{Cooke1978}, and was found to be associated with
a medium compact Zwicky cluster \cite[]{Zwicky1965,Schwartz1980}. 
The presence of a cooling flow was first noted by \citet{Schwartz1980} in the data obtained by \emph{HEAO 1}.
X-ray observations with \emph{EXOSAT} \cite[]{Singh1986,Singh1988} and \emph{Einstein}
\citep{White1991} confirmed the presence of the cooling flow. Observations with \emph{ROSAT} \cite[]{Sarazin1992,Irwin1995} 
show a filamentary structure in the central region of the cooling flow. Observations with \emph{ASCA} \citep{Kikuchi1999} 
show a hint of a centrally peaked metallicity distribution. Using data obtained by \emph{BeppoSAX}, \citet{degrandi2001,degrandi2002}
analyzed the metallicity and temperature profile of the cluster and found a centrally peaked metallicity gradient. 
Using the same dataset, \citet{Ettori2002} estimated the total mass of the cluster within the region with overdensity of 2500 times the critical density to be
$\sim1.6\times10^{14}$ M$_{\odot}$,  while the mass of the gas was found to be $\sim2.0\times10^{13}$ M$_{\odot}$.

Recent \emph{Chandra} observation shows a complex structure in the core of the cluster:
a cold front south of the center, unusual X-ray morphology consisting of a number of X-ray blobs and/or filaments
on scales $\gtrsim3$ kpc, along with two prominent X-ray cavities \citep{Mazzotta2003}. Moreover, the \emph{Chandra} observation shows that the
cluster has a cool dense core and its radial temperature gradient varies with position angle. The radial
metallicity profile has a pronounced central drop and an off-center peak \citep{Mazzotta2003}. A previous shorter observation with \emph{XMM-Newton} 
shows an increase of the Fe abundance toward the center with a strong central peak \citep{tamura2004}.
The central galaxy of 2A~0335+096 is a cD galaxy with a very extended optical emission line region (H$\alpha+$[NII]) to the northeast of 
the galaxy. Moreover, the central region of the
galaxy is anomalously blue, indicating recent star formation \cite[]{Romanishin1988}. \citet[]{edge2001}
reports a detection of CO emission (implying $2\times10^{9}$ M$_{\odot}$ of molecular gas) and IRAS 60 $\mu$m continuum. 
These observations indicate a mass deposition rate of a cooling flow of $<5$ M$_{\odot}$ yr$^{-1}$. 
A radio study of 2A~0335+096 shows a radio source coincident with the central galaxy, which is surrounded by a mini-halo \citep{sarazin1995}. 

Throughout the paper we use $H_{0}=70$ km$\, $s$^{-1}\, $Mpc$^{-1}$, $\Omega_{M}=0.3$, $\Omega_{\Lambda}=0.7$, which imply a linear
scale of 42~kpc\, arcmin$^{-1}$ at the cluster redshift of $z=0.0349$. 
Unless specified otherwise, all errors are at the $1\sigma$ confidence level.

\section{Observations and data reduction}

2A~0335+096 was observed with \emph{XMM-Newton} on August 4th and 5th 2003 with a total exposure of 130 ks. The EPIC MOS and pn instruments 
were operated in Full Frame mode using the thin filter. The exposure times of both MOS cameras and RGS were 130 ks and the exposure time
of pn was 93 ks.

\subsection{EPIC analysis}
The raw data are processed with the 6.0.0 version of the XMM-Newton Science
Analysis System (SAS), using the EPPROC and EMPROC tasks. We apply the standard filtering, for EPIC/MOS
keeping only single, double, triple and quadruple pixel events (PATTERN$\leq12$), while for EPIC/pn we make use of single and double events 
(PATTERN$\leq4$). In both cases only events with FLAG==0 are considered. The redistribution and ancillary response files are created with the SAS tasks {\texttt{rmfgen}} and
{\texttt{arfgen}} for each camera and region we analyze. 

In the spectral analysis we remove all the bright X-ray point-sources with a flux higher than $4.8\times10^{-14}$ erg s$^{-1}$ cm$^{-2}$.

At the low energies, the EPIC cameras are currently not well calibrated below 0.3~keV. For pn we take a conservative lower limit of 0.5~keV, below
which, at the time of our analysis, there are still some uncertainties in calibration. At energies higher than 10~keV our spectra lack sufficient flux. 
Therefore, in most of the analyzed regions, our analysis of the MOS and pn spectra is restricted to the $0.3-10$ keV and $0.5-10$ keV range respectively.
\citet{bonamente2005} showed that the systematic uncertainties of the EPIC detectors in the energy range of 0.3 to 1~keV are less than 10\%. We include
systematic errors to account for the uncertainties in the calibration and background in the spectral fit. The applied systematics are adopted from \citet{kaastra2004}. After
a preliminary analysis we find that the pn instrument has a small gain problem. We minimize this effect by shifting the pn energy grid by 10 eV. The spectra obtained by
MOS1, MOS2 and pn are then fitted simultaneously with the same model, while their relative normalizations are left as free parameters.  

To test whether our abundance measurements can be influenced by narrow band regions where the calibration of EPIC might still be problematic, we fit the
spectrum of PKS~2155-304, a bright BL~Lac object observed with thin filter on 2004 November 23. The X-ray spectra of BL~Lac objects are generally well fitted by
simple absorbed power-law, or broken power-law models. We find that the fit residuals at the position of the measured spectral lines are always consistent within the
$1\sigma$ uncertainty with the best fit absorbed power-law model. Thus, we can rule out the possibility that the measured abundances are seriously influenced by narrow band
calibration uncertainties.

\subsubsection{Background modeling
\label{backgmodels}}

Since we want to investigate the cluster properties up to at least
9$\arcmin$ from its core, we need a good estimate for the background, especially at the dim outer parts of the cluster.
The source fills the entire field of view, which makes a direct measurement of the local background very difficult. The commonly used combined background event lists 
of \citet{lumb2002} and \citet{read2003}, can be used in areas where the surface
brightness of the source is high, or when the cluster has similar background properties as the combined background fields. As shown by \citet{deplaa2005b} relatively small
systematic errors in the background normalization can have a big influence on the spectral fits in the outer parts of clusters. 

The background can be divided into three main components: instrumental background, low energy particles (with energy of a few tens of keV) accelerated in the
Earth's magnetosphere (so called {\it{soft-protons}}) and the Cosmic X-ray Background (CXB). 

To minimize the effect of soft protons in our spectral analysis, we cut out time intervals where the total $10-12$ keV count rate deviates from the 
mean by more than 3$\sigma$. The cleaned MOS1, MOS2 and pn event files have useful exposure times of 107 ks, 108 ks and 77 ks, respectively.

To deal with the instrumental background we adopt a method developed by \citet{deplaa2005b}. As a template for the instrumental background we use data from closed filter
observations, which we scale to the source observation using events detected outside of the field of view (the values of the scaling factors are $1.00\pm0.02$, $1.04\pm0.02$
and $1.10\pm0.07$ for MOS1, MOS2 and pn respectively). The scaling is performed by adding or subtracting a powerlaw with a photon index of $\gamma=0.15$ for MOS and
$\gamma=0.24$ for pn. For detailed description of the method see \citet{deplaa2005b}. 

\begin{table}
\begin{center}
\caption{The CXB components used in the fitting of 2A~0335+096. The power-law index of the extragalactic component is frozen to $\gamma=1.41$. The fluxes are
determined in the $0.3-10$ keV band and, apart of the Local hot bubble, corrected for Galactic absorption using a column of $N_{\mathrm{H}}$=$2.5\times10^{21}$ cm$^{-2}$
(the local hot bubble is inside of the Galactic absorbing column).
\label{tab:back}}
\begin{tabular}{llll}
\hline
\hline
Component		&	k$T$ (keV) 	&	Flux (erg s$^{-1}$ cm$^{-2}$ deg$^{-2}$) \\
\hline
Local bubble		&	0.082 & $3.58\times10^{-12}$ \\
Soft distant component	&	0.068 & $6.57\times10^{-14}$ \\
Hard distant component	&	0.127 & $6.19\times10^{-13}$ \\
Extra galactic power-law &	      & $2.16\times10^{-11}$ \\
\hline
\end{tabular}
\label{tab:cxb}
\end{center}
\end{table}

We correct for the Cosmic X-ray Background (CXB) during the fitting. Since all the CXB photons enter the mirrors together with the source
photons, we can fit the CXB simultaneously with the source spectra. 
In our models, we use CXB components described by \citet{kuntz2000}. They distinguish 4 different components: {\it{the extragalactic power law}} (EPL), {\it{the local hot
bubble}} (LHB), {\it{the soft distant component}} (SDC) and {\it{the hard distant component}} (HDC). 
The EPL component is made from the integrated emission of faint discrete sources, mainly from distant Active Galactic Nuclei (AGNs). According to one of the most recent
determinations by \citet{deluca2004} the power law index of the EPL is $1.41\pm0.06$ and its $2-10$ keV flux is $2.24\pm0.16\times10^{-11}$ erg cm$^{-2}$ s$^{-1}$ deg$^{-2}$
(with 90\% confidence). Since in our spectral analysis we cut out all the point sources with a flux higher than $4.8\times10^{-14}$ erg s$^{-1}$ cm$^{-2}$ we reduce the EPL flux in our
extraction area. \citet{moretti} have made a compilation of number counts of X-ray point sources in two energy bands ($0.5-2$ and $2-10$ keV) from a large source sample and
determined analytic formula for the number of point sources $N(S)$ with a flux higher then $S$. To determine, the contribution of the point sources brighter then
our cut-off to the total flux of the EPL we calculate the integral $\int_{S_{cut-off}}^{\infty} \left( \frac{dN}{dS} \right) S\, dS$. We find, that the point sources we cut
out make up $\approx20$ \% of the total EPL flux. The $0.3-10$ keV flux of the EPL in our model is thus $2.16\times10^{-11}$  erg cm$^{-2}$ s$^{-1}$ deg$^{-2}$.
The LHB is a local supernova remnant, in which our Solar System resides. It produces virtually unabsorbed emission at a temperature of $\sim10^{6}$ K. The SDC and HDC
originate at a larger distance, they might be identified with the Galactic halo, Galactic corona or the Local group emission and are absorbed by almost the full Galactic
column density. They have a temperature of $1-2$ million K. We model each of the three soft background components (the LHB, SDC and the HDC) by a MEKAL model with
temperatures of 0.082 keV, 0.068 keV, 0.127 keV respectively \citep[based on][]{kuntz2000}. We determine the normalizations of these soft components by fitting the spectrum
extracted from an annulus, with inner and outer radii of $9\arcmin$ and $12\arcmin$ respectively, centered at the core of the cluster. The contribution to the total flux
from cluster emission at these radii is $\sim60$\%, the rest of the emission comes from the CXB. To account for the cluster emission we use an additional thermal model. 
In our final spectral fits we fix the normalizations of the background components. The temperatures and the $0.3-10$ keV fluxes of the background components are given in
Table~\ref{tab:back}.

\subsection{RGS analysis}

We extract the RGS spectra with SAS version 6.1.0 following the same method as described in \citet{tamura2001b}.
In order to get spatial information we select the events from several rectangular areas on the CCD strip in the 
cross-dispersion direction. Because the cluster fills the entire field-of-view of the RGS, we need a blank field observation
to extract the background spectrum. For this observation we choose a Lockman Hole observation with an effective exposure time 
of 100 ks.
The flare subtraction is analogous to the method used with EPIC, but now we use the events from CCD 9 
outside the central area with a cross-dispersion of $|xdsp| > 30\arcsec$ to make the lightcurve.
This method was applied to both source and background datasets.

Because the RGS gratings operate without a slit, the resulting spectrum of an extended source is the sum of
all spectra in the (in our case) $5\arcmin$ $\times$ $\sim12\arcmin$ field of view, convolved with the PSF
\citep[see][~for a complete discussion about grating responses]{davis2001}.
Extended line-emission appears to be broadened depending on the spatial extent of the source along the dispersion direction.
In order to describe the data properly, the spectral fits need to account for this effect. In practice, this is accomplished by
convolving the spectral models with the surface brightness profile of the source along the dispersion direction \citep{tamura2004}.
For that purpose we derive for each extraction region the cluster intensity profile from MOS1 in the $0.8-1.4$ keV band along the dispersion direction of RGS and
we convolve this MOS1 profile with the RGS response in order to produce a predicted line spread function (lsf). Because the radial profile for an ion can be different from the
mean MOS1 profile, this method is not ideal. Therefore we introduce two scale parameters, namely the width and centroid of the lsf. These parameters are left free during
spectral fitting in order to match the observed profiles of the main emission lines. The scale parameter $s$ for the width is the ratio of the observed lsf width to the nominal
MOS1 based lsf width.

\section{Spectral models
\label{models}}

For the spectral analysis we use the SPEX package \citep{kaastra1996}. We model the Galactic absorption using the \emph{hot} model of that package, which calculates
the transmission of a plasma in collisional ionisation equilibrium with cosmic abundances. We mimic
the transmission of a neutral plasma by putting its temperature to 0.5 eV. 
To find the best description of the cluster emission we fit several combinations of collisionally ionised equilibrium (CIE) plasma models (MEKAL) to the spectra: a
single-temperature thermal model; a combination of two thermal models; a differential emission measure (DEM) model with a 
cut-off power-law distribution of emission measures versus temperature ({\it{wdem}}). The {\it{wdem}} model appears to be a good empirical approximation for the spectrum
in cooling cores of clusters of galaxies \citep[e.g.][]{kaastra2004,deplaa2004}.  
The emission measure $Y = \int n_{\mathrm{e}} n_{\mathrm{H}} dV$ (where $n_{\mathrm{e}}$ and $n_{\mathrm{H}}$ are the electron and proton densities, $V$ is the volume of the
source) in the {\it{wdem}} model is shown in Eq.~(\ref{eq:wdem}) adapted from \citet{kaastra2004}:
\begin{equation}
\frac{dY}{dT} = \left\{ \begin{array}{ll}
AT^{1/\alpha} & \hspace{1.0cm} T_{\mathrm{min}} < T < T_{\mathrm{max}}, \\
0 & \hspace{1.0cm} \mathrm{elsewhere}. \\
\end{array} \right.
\label{eq:wdem}
\end{equation}

The emission measure distribution has a cut-off at $T_{\mathrm{min}}=cT_{\mathrm{max}}$. The cut-off $c$ is set in this study to 0.1.
For $\alpha \to \infty$ we obtain a flat emission measure distribution. 
The emission measure weighted mean temperature $T_{\mathrm{mean}}$ is given by:
\begin{equation}
T_{\mathrm{mean}}=\frac{\int^{T_{\mathrm{max}}}_{T_{\mathrm{min}}} \frac{dY}{dT}\, T\, dT}{\int^{T_{\mathrm{max}}}_{T_{\mathrm{min}}} \frac{dY}{dT}dT}.
\end{equation}
By integrating this equation between $T_{\mathrm{min}}$ and $T_{\mathrm{max}}$ we obtain a direct relation between $T_{\mathrm{mean}}$ and $T_{\mathrm{max}}$ as a function of
$\alpha$ and $c$:
\begin{equation}
T_{\mathrm{mean}}=\frac{(1+1/\alpha)(1-c^{1/\alpha+2})}{(2+1/\alpha)(1-c^{1/\alpha+1})}T_\mathrm{max}.
\end{equation}
A comparison of the \emph{wdem} model with the classical cooling-flow model can be found in \citet{deplaa2005a}. We note that the \emph{wdem} model contains less cool gas than
the classical cooling-flow model, which is consistent with recent observations \citep[]{peterson2001,peterson2003}.

The spectral lines in the MEKAL model are fitted self consistently. 
From the fits we obtain the numbers of atoms of all elements with detected line emission. To see whether
different elements show similar abundances with respect to solar, it is convenient to normalize these
numbers with respect to the relative solar abundances. To make the comparison with previous work easier, we use the solar abundances as
given by \citet{anders1989} for this normalization. 
The more recent solar abundance determinations \cite[e.g.][]{Grevesse1998,lodders2003} give significantly lower abundances of oxygen and neon than those measured by
\citet{anders1989}. Use of these new determinations would only affect the {\it representation} of the elemental abundances in our paper, but not the actual measured
values, which can be reconstructed by multiplication of the given values with the normalizations.

\section{Global spectrum}
\subsection{EPIC}
\begin{figure}
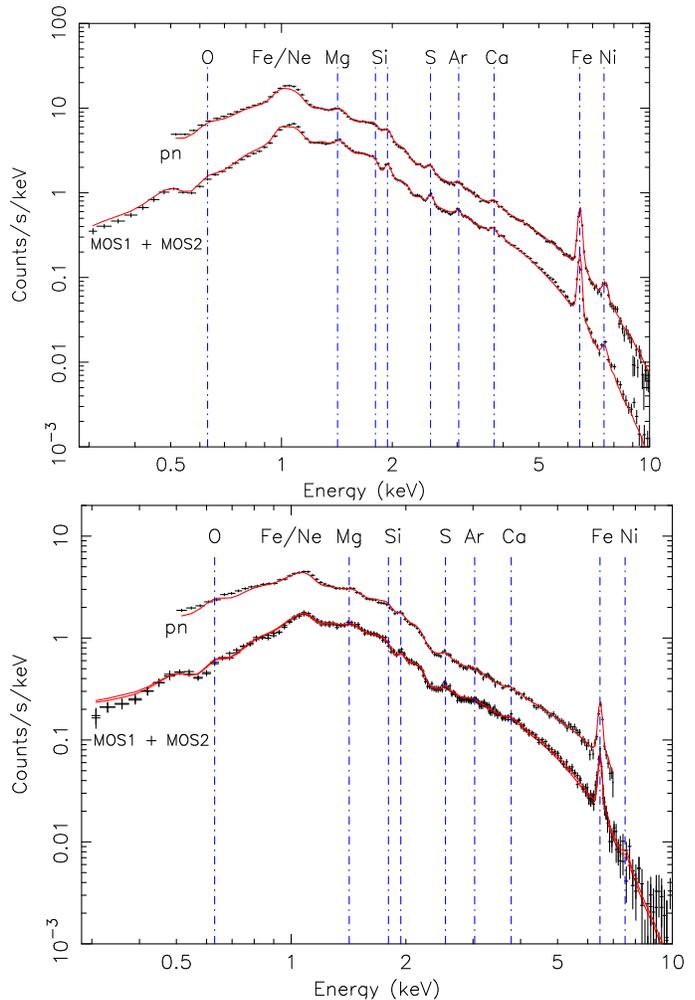

\begin{minipage}{\columnwidth}
\includegraphics[width=0.75\textwidth,clip=t,angle=270.]{3868f1a.ps}
\end{minipage}
\begin{minipage}{\columnwidth}
\includegraphics[width=0.75\textwidth,clip=t,angle=270.]{3868f1b.ps}
\end{minipage}
\caption{The total spectrum of the core of the cluster (with a radius of $3\arcmin$), which contains the cooling core (upper panel) and the spectrum of the
$3\arcmin-9\arcmin$  region (lower panel). The continuous line represents the fitted \emph{wdem} model. The EPIC pn and MOS1 + MOS2 spectra are indicated.}
\label{fig:globalspectr}
\end{figure}

\begin{table}
\begin{center}
\caption[]{Fit results obtained by fitting the {\it{wdem}} model to the spectrum extracted from a circular region with a radius of
$3\arcmin $ centered on the core of the cluster and from an annulus with inner and outer radii respectively $3\arcmin $ and $9\arcmin $. For the Galactic absorption we use a
value of $N_{\mathrm{H}}$=$2.5\times10^{21}$ cm$^{-2}$. Emission measures ($Y = \int n_{\mathrm{e}} n_{\mathrm{H}} dV$) are given in $10^{66}$ cm$^{-3}$. Abundances are given
with respect to solar.} 

\label{global}
\begin{tabular}{l|cc}
\hline
\hline
Parameter & $0-3\arcmin $ & $3-9\arcmin $ \\
\hline
$Y$ & $15.89\pm0.13 $ & $6.53\pm0.01$ \\
$kT_{\mathrm{max}}$ (keV)   & $3.75\pm0.02$   & $5.36\pm0.06$ \\
$\alpha$   &          $0.75\pm0.02$   & $1.34\pm0.10$ \\
$kT_{\mathrm{mean}}$  & $2.64\pm0.02$   & $3.46\pm0.05$\\
Mg                    & $0.57\pm0.08$   & $0.21\pm0.06$ \\    
Si                    & $0.72\pm0.03$   & $0.37\pm0.03$ \\    
S                     & $0.61\pm0.03$   & $0.27\pm0.04$\\     
Ar                    & $0.42\pm0.05$   & $0.14\pm0.09$\\     
Ca                    & $0.84\pm0.06$   & $0.62\pm0.12$\\     
Fe                    & $0.532\pm0.007$ & $0.383\pm0.006$\\      
Ni                    & $1.46\pm0.10$   & $0.77\pm0.15$\\    
\hline 
$\chi^{2}$ / d.o.f. & 948/525 & 666/491 \\ 
\hline
\end{tabular}
\end{center}

\end{table}

\begin{table}
\begin{center}
\caption{Abundance upper limits for elements with weak lines, which could not be reliably detected, determined from the fluxes
at the expected line energies of their helium-like emission. We also show the abundance of calcium calculated with the same method (it is consistent with the abundance
determined in MEKAL).}
\label{cr}
\begin{tabular}{l|cc}
\hline
\hline
Element & I (phot m$^{-2}$ s$^{-1}$) & Abund. (solar)\\
\hline
Ca & $0.252\pm0.019$ & $0.80\pm0.05$\\
Ti & $0.017\pm0.013$ & $1.5\pm1.1$\\
Cr & $0.020\pm0.010$ & $0.5\pm0.2$\\
Mn & $-0.003\pm0.009$ & $-0.2\pm0.5$\\
Co & $0.000\pm0.008$ & $0\pm2$\\
\hline

\end{tabular}
\end{center}

\end{table}

\begin{figure}
\psfig{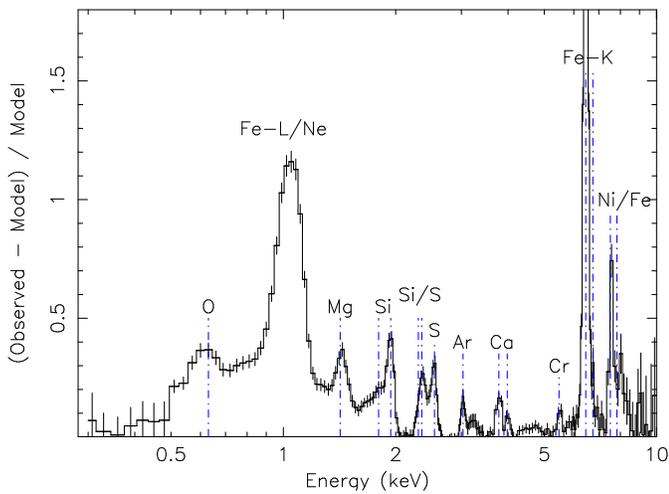}
\caption{Residuals of the fit to the EPIC total spectrum extracted from a circular region with a radius of $3\arcmin$, with the line emission put to zero in the model.}
\label{lines}
\end{figure}

To compare the properties of the cluster in its cooling core region with the properties outside of the cooling core region we extract two spectra.
One is from a circular region with a radius of $3\arcmin$ centered on the X-ray maximum of the cluster and one is from an annulus with inner radius of $3\arcmin$
and outer radius of $9\arcmin$. The large temperature gradient and projection effects make it necessary to fit the spectra with a multi-temperature model. 
We use the {\it{wdem}} model, described in Sect.~\ref{models} and fix the abundances of elements with weak lines, which we do not fit, to 0.6 and 0.3 times solar in
the inner and outer region respectively. The high statistics of the spectra allow us to determine the temperature structure
and elemental abundances of several elements in these regions very precisely. The best-fit parameters are shown in Table~\ref{global}. 

We find that the core is cooler than the outer region. We also find that the outer part has a broader temperature distribution than the core. 
The spectra with the indicated important spectral lines are shown in Fig.~\ref{fig:globalspectr}. 
All the detected spectral lines are shown in Fig.~\ref{lines}, which shows the residuals of the EPIC spectrum with line emission put to zero in the model, so
all spectral lines are clearly visible. For the first time we see a feature at the expected energy of chromium in a cluster of galaxies, which corresponds to a detection with
$2\sigma$ significance. First, we attempt to measure the abundances of elements shown in Table~\ref{global}. 

Since oxygen is predominantly produced by type II supernovae, it is an important element for constraining the enrichment scenarios. Oxygen has strong H-like
lines at 0.65 keV. The strong absorption toward 2A~0335+096 (see Appendix~\ref{infback}), uncertainties in the calibration of EPIC and relatively
nearby iron lines do not allow us to constrain the oxygen abundance accurately by EPIC. 
However, the RGS with its high spectral resolution allows us to determine the oxygen abundance in the core of the cluster (see
Table~\ref{tab:rgs_240}). The 2p$-1$s neon lines at 1.02 keV are in the middle of the iron L complex (lying between about 0.8 to 1.4 keV). The resolution of the EPIC
cameras is not sufficient to resolve the individual lines in the iron L complex which makes the neon abundance determination by EPIC unreliable. However, the high
resolution of RGS allows us to determine the neon abundance at least in the core of the cluster (see Table~\ref{tab:rgs_240}). 
The K shell lines of magnesium at 1.47 keV are also close to the iron L complex, which makes our magnesium abundance determinations by EPIC somewhat sensitive to our
temperature and iron abundance model. The magnesium abundance determinations at larger radii, where the surface brightness of the cluster is relatively low might be
influenced by the instrumental aluminum line at 1.48 keV.  
The silicon, sulfur, argon and calcium lines lie in a relatively uncrowded part of the spectrum and their abundances are in general well determined.
Iron has the strongest spectral lines in the X-ray band.
At temperatures above 3 keV its K$\alpha$ lines are the strongest at about 6.67 keV and 6.97 keV, while the iron L-shell complex ranging from about 0.8 keV to 1.4
keV dominates at lower temperatures. These spectral lines make the iron abundance determinations the most reliable of all elements. The nickel abundance is determined mainly
from its K-shell line blends at 7.80 keV and 8.21 keV, which are partially blended with iron lines.

In Table~\ref{global} we see that the abundances in the outer part are always lower than in the core of the cluster. 

We also attempt to estimate the abundance upper limits for elements with weak lines (titanium, chromium, manganese, cobalt), which can not yet be fitted in MEKAL in a
self-consistent way. We determine their abundances from the fluxes at the expected line energies of their helium-like emission (see Table~\ref{cr}). The feature at the
expected energy of chromium corresponds to a detection at a $2\sigma$ level, while the feature at the expected energy of titanium corresponds to a $1\sigma$ detection.

\subsection{RGS}

\label{RGSres}
\begin{figure}
\sidecaption
\includegraphics[width=0.74\columnwidth,angle=270]{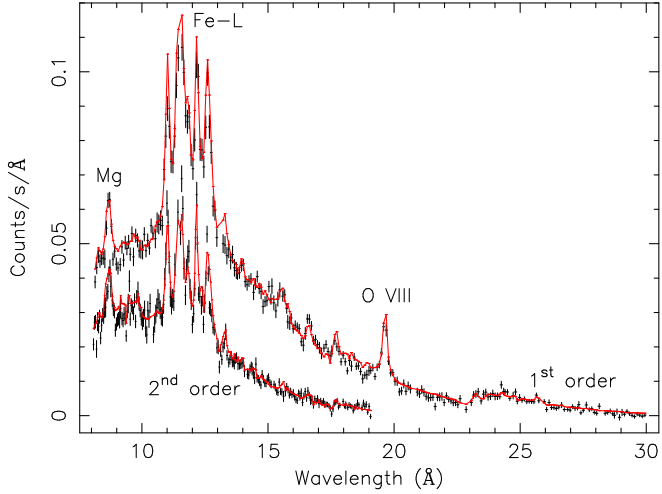}
\caption{$~1^{st}$ and 2$^{nd}$ order spectrum of 2A~0335+096 extracted from a 4$\arcmin$ wide strip 
centered on the core. The continuous line represents the fitted {\it{wdem}} model. Between 15 and 18 $\AA$ Fe~XVII and Fe~XVIII emission lines are visible. On the x-axis
we show the observed wavelength. }
\label{fig:rgs_wdem}
\end{figure}

\begin{table}
\caption{Fit results for the RGS spectra extracted from a 4$\arcmin$ wide strip. 
The value for $N_{\mathrm{H}}$ is in 10$^{21}$ cm$^{-2}$. Iron is given with respect to solar 
and the other abundances with respect to iron.}
\begin{center}
\begin{tabular}{l|ccc}
\hline\hline
Parameter 	& single-temp	 	& {\it wdem}-model 	 \\
\hline
$N_{\mathrm{H}}$& 3.14 $\pm$ 0.03	& 2.95 $\pm$ 0.04        \\
$kT$ (keV)	& 1.80 $\pm$ 0.02	&   	 		 \\
$kT_{\mathrm{max}}$ & 			& 3.36 $\pm$ 0.13	\\
$kT_{\mathrm{mean}}$ & 			& 2.51 $\pm$ 0.10	\\
$\alpha$	& 			& 0.52 $\pm$ 0.03 	\\
O$_{\mathrm{abs}}$ & 0.47 $\pm$ 0.02	& 0.54 $\pm$ 0.03	     \\
N/Fe		& 1.7 $\pm$ 0.6		& 1.3  $\pm$ 0.4	 \\  
O/Fe 		& 0.55 $\pm$ 0.05	& 0.49 $\pm$ 0.05	  \\
Ne/Fe		& 1.11 $\pm$ 0.10	& 0.85 $\pm$ 0.08	   \\
Mg/Fe		& 0.97 $\pm$ 0.10	& 0.97 $\pm$ 0.08	   \\
Fe		& 0.524 $\pm$ 0.018	& 1.07 $\pm$ 0.06	   \\
Scale $s_{O}$ 	& 0.59 $\pm$ 0.12	& 0.81 $\pm$ 0.14	  \\
Scale $s_{Fe}$	& 0.60 $\pm$ 0.04	& 0.73 $\pm$ 0.04 	   \\
\hline
$\chi^2$ / d.o.f. & 1552 / 790		& 1173 / 788		\\
\hline
\end{tabular}
\label{tab:rgs_240}
\end{center}
\end{table}

In Fig.~\ref{fig:rgs_wdem} we show the first and second order RGS spectra extracted from a 4$\arcmin$ wide strip in the cross-dispersion direction of
the instrument. The extraction region is centered on the core of the cluster. Despite the spatial broadening, the strong spectral lines of magnesium,
neon, iron and oxygen are well resolved in both spectral orders. In Table~\ref{tab:rgs_240} we list the fit results for these spectra, for both the single temperature
and {\it{wdem}} model.

The $\chi^2$ values of the fits show that the {\it wdem} model fits better than the single-temperature model.
However, the $\chi^2$ for the {\it wdem} is still high, probably because of the disturbed nature of the core of the cluster. 
There are some systematic differences between the two models. 
The abundances found fitting the spectrum with the {\it wdem} model are in general twice the single-temperature values and the line widths of oxygen and iron,
indicated with the scale parameters $s_{O}$ and $s_{Fe}$, are also higher for {\it wdem}.
Because of these differences, we divide the abundances of the other elements, by the value for iron. Since the continuum is not well determined by the RGS, these
ratios are more reliable than the absolute value. The relative abundances, given in units with
respect to solar, for neon and magnesium are similar to iron, while the oxygen abundance is about half the value for iron. We are also able, using the excellent statistics, to
derive the abundance for nitrogen. This value is also consistent with that for iron, although the error bars are large. 
The widths of the oxygen and iron lines, indicated by the scale values,
are very similar in this cluster, which shows that the spatial distribution of oxygen and iron might be similar. 

Regarding the DEM temperature structure, the value of $\alpha$ (0.52 $\pm$ 0.03) for {\it wdem} is smaller than the EPIC value
of 0.75 $\pm$ 0.02 from within 3$\arcmin$ from the core, probably, because of the different extraction regions. 
The different extraction regions and the disturbed nature of the core (see sections \ref{maps} and \ref{smallscale})
make the abundances determined from the EPIC and RGS difficult to compare.
The mean temperature $kT_\mathrm{mean}$ of the {\it wdem} results is 2.51 $\pm$ 0.10 keV, which is higher than the single-temperature fit.
 
Because the solar oxygen abundance in \citet{anders1989}
is slightly overestimated, we need to free the oxygen-abundance in the Galactic absorption component (O$_{\mathrm{abs}}$) in our spectral model. This effect 
in absorption was first observed by \citet{weisskopf2004} in an observation of the Crab pulsar and later confirmed in a cluster observation of 
\object{Abell 478} by \citet{deplaa2004}. The values of 0.47 $\pm$ 0.02 and 0.54 $\pm$ 0.03 that we find for the oxygen-abundance in the Galactic absorption component
O$_{\mathrm{abs}}$ are slightly lower than the RGS values found by \citet{deplaa2004}, but consistent with the expected value of 0.58 $\pm$ 0.08 based on an updated value for
the solar oxygen abundance by \citet{allende2001}.

\section{Radial profiles}
\label{radial}

We determine the radial temperature and abundance profiles of several elements using projected and deprojected spectra
extracted from circular annuli, centered on the X-ray maximum of the cluster. We use annuli with outer radii indicated on the top of Table \ref{3models}. We also extract radial
temperature and abundance profiles from RGS in the cross-dispersion direction.

\subsection{Projected spectra
\label{modelcomp}}

\begin{figure*}
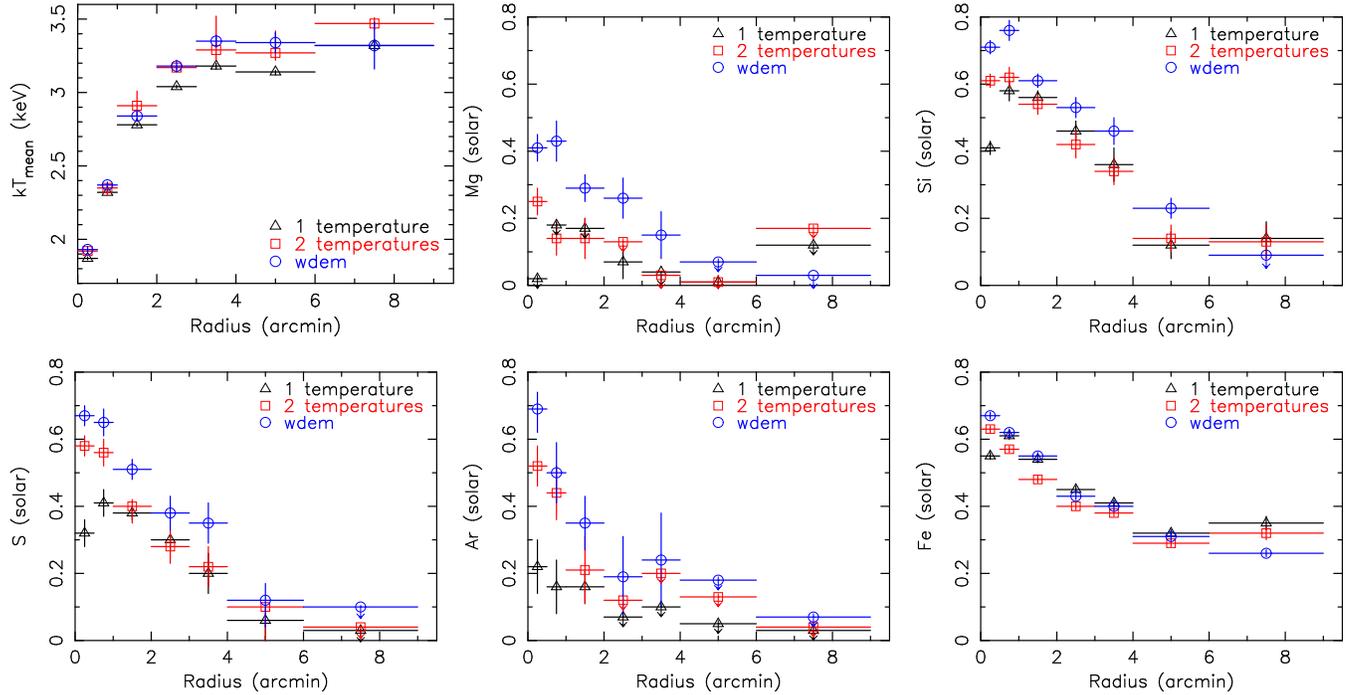

\begin{minipage}{0.33\textwidth}
\includegraphics[width=0.75\textwidth,clip=t,angle=270.]{3868f4a.ps}
\end{minipage}
\begin{minipage}{0.33\textwidth}
\includegraphics[width=0.75\textwidth,clip=t,angle=270.]{3868f4b.ps}
\end{minipage}
\begin{minipage}{0.33\textwidth}
\includegraphics[width=0.75\textwidth,clip=t,angle=270.]{3868f4c.ps}
\end{minipage}\\
\vspace{2mm}

\begin{minipage}{0.33\textwidth}
\includegraphics[width=0.75\textwidth,clip=t,angle=270.]{3868f4d.ps}
\end{minipage}
\begin{minipage}{0.33\textwidth}
\includegraphics[width=0.75\textwidth,clip=t,angle=270.]{3868f4e.ps}
\end{minipage}
\begin{minipage}{0.33\textwidth}
\includegraphics[width=0.75\textwidth,clip=t,angle=270.]{3868f4f.ps}
\end{minipage}\\
\caption{Comparison of fit results obtained by fitting the EPIC data with a single-temperature model, two-temperature model and the
{\it{wdem}} model. } 
\label{fig:3models}
\end{figure*}

\begin{table*}
\begin{center}
\caption[]{Fit results obtained by fitting the EPIC data with a single-temperature model (1), two-temperature model (2) and the
{\it{wdem}} model (3). Emission measures ($Y = \int n_{\mathrm{e}} n_{\mathrm{H}} dV$) are given in $10^{66}$ cm$^{-3}$. Abundances are given with respect to solar.}
\label{3models}
\begin{tabular}{l|c|ccccccc}
\hline
\hline
 & Model  & $0-0.5\arcmin $ & $0.5-1.0\arcmin $  & $1.0-2.0\arcmin $ & $2.0-3.0\arcmin $ & $3.0-4.0\arcmin $ & $4.0-6.0\arcmin $ & $6.0-9.0\arcmin $\\
\hline
Y			      & 1 & 3.55 & 4.65 & 5.22 & 3.19 & 2.01 & 2.85 & 2.97 \\
			      & 2 & 3.42 & 4.82 & 5.45 & 3.31 & 2.08 & 2.75 & 2.56\\
			      & 3 & 3.21 & 4.52 & 5.30 & 3.26 & 2.05 & 2.85 & 2.23 \\
$kT$ (keV)                    & 1 & $1.87\pm0.01$ & $2.32\pm0.01$ & $2.78\pm0.01$ & $3.04\pm0.01$ & $3.18\pm0.02$ & $3.14\pm0.02$ & $3.32\pm0.04$\\
$kT_{\mathrm{2}}$ (keV)  	      & 2 & $2.154^{+0.005}_{-0.002}$ & $2.72\pm0.04$ & $3.51\pm0.09$ & $3.08\pm0.01$ & $3.92^{+0.23}_{-0.07}$ & $4.10^{+0.15}_{-0.05}$ & $4.39\pm0.04$\\
$kT_{\mathrm{max}}$ (keV)              & 3 & $2.64\pm0.02$ & $3.14\pm0.02$ & $3.95\pm0.03$ & $4.39\pm0.04$ & $4.81\pm0.05$ & $4.97\pm0.05$ & $5.84\pm0.14$ \\
$kT_{\mathrm{mean}}$ (keV)             & 3 & $1.93\pm0.02$ & $2.37\pm0.02$ & $2.84\pm0.03$ & $3.18\pm0.04$ & $3.35\pm0.04$ & $3.34\pm0.08$ & $3.32\pm0.16$ \\
$\alpha$                      & 3 & $0.58\pm0.01$ & $0.48\pm0.02$ & $0.65\pm0.02$ & $0.62\pm0.04$ & $0.79\pm0.04$ & $1.01^{+0.23}_{-0.02}$ &$8.02^{+13}_{-3}$ \\
\hline
Mg                            & 1 & $<0.02$ & $<0.18$ & $0.17\pm0.03$ & $0.07\pm0.05$ & $<0.04$ & $<0.01$ & $<0.12$ \\
	   	              & 2 & $0.25\pm0.04$ & $0.14\pm0.05$ & $0.14\pm0.06$ & $<0.13$ & $<0.03$ & $<0.01$ & $<0.17$ \\
                              & 3 & $0.41\pm0.04$ & $0.43\pm0.06$ & $0.29\pm0.04$ & $0.26\pm0.06$ & $0.15\pm0.07$ & $<0.07$ & $<0.03$\\
\hline
Si                            & 1 & $0.41\pm0.02$ & $0.58\pm0.03$ & $0.56\pm0.02$ & $0.46\pm0.03$ & $0.36\pm0.05$ & $0.12\pm0.04$ & $0.14\pm0.05$ \\
	   	              & 2 & $0.61\pm0.02$ & $0.62\pm0.03$ & $0.54\pm0.03$ & $0.42\pm0.04$ & $0.34\pm0.04$ & $0.14\pm0.04$ & $0.13\pm0.05$\\
                              & 3 & $0.71\pm0.02$ & $0.76\pm0.03$ & $0.61\pm0.02$ & $0.53\pm0.03$ & $0.46\pm0.04$ & $0.23\pm0.03$ & $<0.09$ \\
\hline
S                             & 1 & $0.32\pm0.04$ & $0.41\pm0.04$ & $0.38\pm0.02$ & $0.30\pm0.02$ & $0.20\pm0.06$ & $<0.06$ & $<0.03$ \\
	   	              & 2 & $0.58\pm0.03$ & $0.56\pm0.04$ & $0.40\pm0.04$ & $0.28\pm0.05$ & $0.22\pm0.06$ & $<0.10$ & $<0.04$ \\
                              & 3 & $0.67\pm0.03$ & $0.65\pm0.04$ & $0.51\pm0.03$ & $0.38\pm0.05$ & $0.35\pm0.06$ & $0.12\pm0.07$ & $<0.10$ \\
\hline
Ar                            & 1 & $0.22\pm0.08$ & $0.16\pm0.08$ & $0.16\pm0.05$ & $<0.07$ & $<0.10$ & $<0.05$ & $<0.03$ \\
	   	              & 2 & $0.52\pm0.06$ & $0.44\pm0.08$ & $0.21\pm0.10$ & $<0.12$ & $<0.20$ & $<0.13$ & $<0.04$ \\
                              & 3 & $0.69\pm0.06$ & $0.50\pm0.09$ & $0.35\pm0.08$ & $0.19\pm0.12$& $0.24\pm0.14$ & $<0.18$ & $<0.07$ \\		      
\hline
Ca                            & 1 & $0.51\pm0.12$ & $0.66\pm0.10$ & $0.85\pm0.07$ & $0.68\pm0.09$ & $0.53\pm0.15$ & $0.57\pm0.15$ & $0.27\pm0.21$ \\
	   	              & 2 & $0.66\pm0.09$ & $0.96\pm0.10$ & $0.92\pm0.12$ & $0.71\pm0.16$ & $0.66\pm0.14$ & $0.70\pm0.14$ & $0.30\pm0.23$ \\
                              & 3 & $0.74\pm0.09$ & $0.98\pm0.11$ & $1.02\pm0.10$ & $0.77\pm0.15$ & $0.74\pm0.17$ & $0.66\pm0.17$ & $0.70\pm0.23$ \\		      
\hline
Fe                            & 1 & $0.55\pm0.01$ & $0.61\pm0.01$ & $0.54\pm0.01$ & $0.45\pm0.01$ & $0.41\pm0.01$ & $0.32\pm0.01$ & $0.35\pm0.02$ \\
	   	              & 2 & $0.63\pm0.01$ & $0.57\pm0.01$ & $0.48\pm0.01$ & $0.40\pm0.01$ & $0.38\pm0.01$ & $0.29\pm0.01$ & $0.32\pm0.02$\\
                              & 3 & $0.67\pm0.01$ & $0.62\pm0.01$ & $0.55\pm0.01$ & $0.43\pm0.01$ & $0.40\pm0.01$ & $0.31\pm0.01$ & $0.26\pm0.01$ \\			      
\hline
Ni                            & 1 & $0.68\pm0.11$ & $1.08\pm0.13$ & $0.57\pm0.08$ & $0.35\pm0.11$ & $0.37\pm0.20$ & un. & un.\\
	   	              & 2 & $0.89\pm0.09$ & $0.90\pm0.12$ & $0.53\pm0.13$ & $0.29\pm0.15$ & $0.22\pm0.17$ & un. & un.\\
                              & 3 & $1.65\pm0.10$ & $1.52\pm0.14$ & $0.99\pm0.10$ & $0.64\pm0.20$ & $0.69\pm0.17$ & un. & un.\\			  
\hline
$\chi^{2}$ / d.o.f.           & 1 & 2444/525 & 1278/525 & 1066/525 & 774/525 & 717/525 & 909/499 & 816/476 \\
	                      & 2 & 844/525 & 753/525 & 883/525 & 680/525 & 664/525 & 789/499 & 742/476 \\
                              & 3 & 782/525 & 664/525 & 707/525 & 596/525 & 532/525 & 607/499 & 880/476 \\
\hline

\end{tabular}
\noindent
\end{center}
\end{table*}

We fit the spectra extracted from the annuli with three different models, and compare the results. Due to high background in the 6th and 7th annulus we ignore the pn data at
energies higher than 7.5 keV. In our model we fix the abundances of elements which we do not fit to 0.3 solar. The results are presented in
Fig.~\ref{fig:3models} and  Table~\ref{3models}.  First we fit the spectra with a single temperature model. We find that the model fits our data poorly and the $\chi^{2}$
in the core of the cluster is  unacceptably high. Based on the large $\chi^2$ of the fit, in the central region we can discard the single temperature model. 
We therefore try to fit 2 thermal models simultaneously, with coupled abundances and with a fixed separation between the two temperatures as
$T_{2}=2T_{1}$ (where $T_{1}$ and $T_{2}$ are the temperatures of the two thermal components). Fixing the temperature to $T_{2}=2T_{1}$ is a good first approximation of the
multi-temperature structure, since \citet{kaastra2004} and \citet{peterson2003} found that in almost all cases of their cluster samples,
at each radius, there is negligible emission from gas with a temperature less than one-third to half of the fitted upper temperature. Using the two temperature
model our fits improve significantly. The $\chi^{2}$ improves most significantly in the core of the cluster, but we can see a slight improvement also at outer radii. 
However, fitting the data with the {\it{wdem}} model further improves the $\chi^{2}$ of our fits in 6 of 7 extraction annuli. 

Fitting the data with a single-temperature thermal model we see that the abundances of all elements except argon drop in the core of the cluster and have a off-center peak. 
As we go from the single-temperature model to the two-temperature model and further to the {\it{wdem}} model, abundances of all elements in general increase,
especially in the central regions of the cluster. The {\it{wdem}} model shows that most of the abundances peak in the core of the cluster, we detect a slight hint of a drop in
the center only for calcium, silicon and magnesium. The dependence of the iron abundance on the temperature model is known as the Fe-bias \citep[see,
e.g.][]{Buote2000,Molendi2001}. As we go from one model to the other, the abundance value of magnesium varies the most. Due to its vicinity to the Fe-L complex its abundance
is very sensitive to the temperature and iron abundance model. 

In general, all three models show a drop of abundances toward the outer parts of the cluster.

\subsection{Deprojected spectra}

To account for projection effects in the core of the cluster we extract spectra, deprojected under the assumption of a spherical symmetry. We use the background event
files of \citet{lumb2002}. We extract the deprojected spectra from 5 annuli in the inner $0\arcmin-4\arcmin$, where the cluster is still bright enough and, as shown in
Appendix \ref{infback}, the background subtraction does not have a significant influence on our results. 
The deprojected spectra represent the count rates from spherical shells centered on the core of the cluster. Our extraction method and data analysis for the deprojected
spectra is described extensively in \citet{kaastra2004}. 

\begin{figure}
\psfig{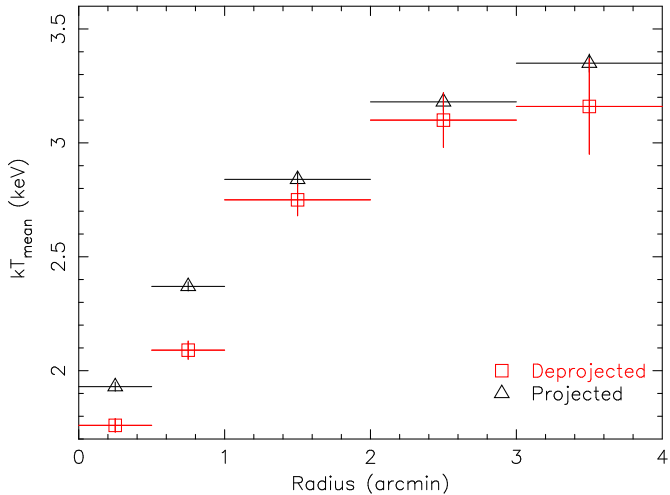}
\caption{Comparison of the projected and deprojected mean temperatures obtained by fitting the {\it{wdem}} model.}
\label{DvP}
\end{figure}

The fit results for the deprojected spectra are shown in Table~\ref{tab:deproject}. The temperatures determined from deprojected and projected spectra are compared in
Fig.~\ref{DvP}. Since, in the process of deprojection we account for the hot plasma lying in front of the relatively cool core, the observed temperatures of the
deprojected shells are lower than those determined without deprojection. While the $\chi^{2}$ of the fit of the single-temperature model improves significantly if we
deproject the spectrum (compare with the $\chi^{2}$ in Table~\ref{3models}), the single temperature thermal model still does not describe our spectra well.
The {\it{wdem}} model describes the deprojected spectra extracted from the core of the cluster significantly better than the single-temperature model. 
It shows, that the temperature gradients in the extraction regions and the intrinsic multi-temperature structure make it necessary to use multi-temperature models also when
fitting deprojected spectra.
Although the deprojected spectra give a better value for the temperature, with the deprojection we introduce more noise into the spectra so the determination of abundances
becomes more uncertain.

\begin{table*}[tb]
\begin{center}
\caption[]{Fit results of the deprojected EPIC spectra extracted from 5 annuli in the inner $4\arcmin$ and fitted with a single-temperature model (1) and {\it{wdem}}
model (3). This region contains the cooling core, where the projection effects are the most important. Abundances are given  with respect to solar.}
\label{tab:deproject}
\begin{tabular}{l|c|ccccc}
\hline
\hline
Parameter & Model  & $0-0.5\arcmin $ & $0.5-1.0\arcmin $  & $1.0-2.0\arcmin $ & $2.0-3.0\arcmin $ & $3.0-4.0\arcmin $ \\
\hline
$N_{\mathrm{H}}$ ($10^{21} $cm$^{-2}$) & 1 & $2.36\pm0.02$ & $2.37\pm0.02$ & $2.47\pm0.03$ & $2.48\pm0.05$ & $2.53\pm0.05$ \\
	                      & 3 & $2.52\pm0.06$ & $2.58\pm0.05$ & $2.64\pm0.08$ & $2.64\pm0.11$ & $2.66\pm0.16$ \\   
\hline
$kT$ (keV)                    & 1 & $1.65\pm0.02$ & $2.07\pm0.02$ & $2.68\pm0.02$ & $3.04\pm0.03$ & $3.11\pm0.05$  \\
$kT_{\mathrm{max}}$ (keV)              & 3 & $2.36\pm0.04$ & $2.72\pm0.04$ & $3.65\pm0.08$ & $4.28\pm0.13$ & $4.25\pm0.21$ \\
$kT_{\mathrm{mean}}$ (keV)             & 3 & $1.76\pm0.03$ & $2.09\pm0.04$ & $2.75\pm0.07$ & $3.10\pm0.12$ & $3.16\pm0.21$  \\
\hline
$\alpha$                      & 3 & $0.52\pm0.03$ & $0.43\pm0.03$ & $0.49\pm0.04$ & $0.62\pm0.08$ & $0.53\pm0.14$ \\
                              & 3 & $0.6\pm0.2$ & $0.7\pm0.2$ & $0.3\pm0.2$ & $0.3\pm0.2$ & $<0.5$\\
\hline
Mg                            & 1 & $0.11\pm0.08$ & $0.34\pm0.10$ & $0.65\pm0.07$ & $0.35\pm0.13$ & $<0.16$ \\
                              & 3 & $0.55\pm0.10$ & $0.53\pm0.11$ & $0.76\pm0.14$ & $0.36\pm0.15$ & $<0.26$ \\
\hline
Si                            & 1 & $0.46\pm0.04$ & $0.73\pm0.05$ & $0.80\pm0.04$ & $0.73\pm0.07$ & $0.52\pm0.09$ \\
                              & 3 & $0.74\pm0.06$ & $0.84\pm0.06$ & $0.84\pm0.07$ & $0.71\pm0.08$ & $0.50\pm0.10$  \\
\hline
S                             & 1 & $0.42\pm0.04$ & $0.61\pm0.05$ & $0.52\pm0.04$ & $0.47\pm0.08$ & $0.19\pm0.12$ \\
                              & 3 & $0.66\pm0.06$ & $0.75\pm0.06$ & $0.61\pm0.07$ & $0.51\pm0.09$ & $0.25\pm0.10$ \\
 \hline
Ar                            & 1 & $0.40\pm0.11$ & $0.50\pm0.11$ & $0.25\pm0.10$ & $<0.14$ & $0.3\pm0.2$\\
                              & 3 & $0.57\pm0.14$ & $0.69\pm0.12$ & $0.33\pm0.15$ & $<0.21$ & $0.4\pm0.3$\\
\hline
Ca                            & 1 & $0.64\pm0.20$ & $0.81\pm0.18$ & $1.00\pm0.13$ & $0.9\pm0.2$ & $0.4\pm0.3$ \\
                              & 3 & $0.41\pm0.20$ & $0.88\pm0.16$ & $1.06\pm0.18$ & $1.0\pm0.3$ & $0.5\pm0.4$ \\
\hline
Fe                            & 1 & $0.50\pm0.02$ & $0.65\pm0.02$ & $0.61\pm0.02$ & $0.46\pm0.02$ & $0.41\pm0.03$  \\
                              & 3 & $0.67\pm0.03$ & $0.66\pm0.03$ & $0.57\pm0.02$ & $0.45\pm0.02$ & $0.40\pm0.03$ \\
\hline
Ni                            & 1 & $0.62\pm0.17$ & $1.8\pm0.2$ & $1.53\pm0.18$ & $0.9\pm0.4$ & $0.6\pm0.5$ \\
                              & 3 & $1.3\pm0.3$ & $2.0\pm0.3$ & $1.8\pm0.3$ & $0.8\pm0.4$ & $0.6\pm0.5$ \\
\hline
$\chi^{2}$ / d.o.f.           & 1 & 1090/587 & 668/570 & 892/587 & 515/587 & 520/587 \\
                              & 3 & 622/587 & 489/587 & 526/587 & 481/587 & 515/587  \\ 
\hline

\end{tabular}
\noindent
\end{center}
\end{table*}

\subsection{RGS radial profiles}

\begin{table*}
\caption{Fit results for spatially resolved RGS spectra between 8--25 \AA. The fitted models are (1) 
single-temperature CIE and (3) {\it wdem}. We use 2$\sigma$ upper limits.}
\begin{center}
\begin{tabular}{l|c|cccccc}
\hline\hline
Parameter	& Model	& -2.0 / -1.0$\arcmin$  & -1.0 / -0.5 $\arcmin$ & -0.5 / 0$\arcmin$ 	& 0 / 0.5$\arcmin$	& 0.5--1.0$\arcmin$	& 1.0--2.0$\arcmin$ \\
\hline
$kT$ (keV)	& 1	& 4.12 $\pm$ 0.13	& 3.37 $\pm$ 0.10	& 2.89 $\pm$ 0.06 	& 2.53 $\pm$ 0.05 	& 3.21 $\pm$ 0.12 	& 4.4 $\pm$ 0.3  \\
$kT_{\mathrm{max}}$ & 3	& 6.6 $\pm$ 0.3		& 5.2 $\pm$ 0.2 	& 4.6 $\pm$ 0.2  	& 3.73 $\pm$ 0.14	& 5.3 $\pm$ 0.8		& 7.8 $\pm$ 0.9 \\
$kT_{\mathrm{mean}}$ & 3 & $4.6\pm0.2$ & $3.70\pm0.15$ & $3.12\pm0.15$ & $2.72\pm0.11$ & $3.8\pm0.7$ & $5.1\pm0.6$\\
\hline
$\alpha$	& 3	& 0.79 $\pm$ 0.08	& 0.70 $\pm$ 0.06	& 0.94 $\pm$ 0.10 	& 0.60  $\pm$ 0.05	& 0.7 $\pm$ 0.4		& 1.3 $\pm$ 0.3 \\
\hline
O/Fe		& 1	& 0.28 $\pm$ 0.07	& 0.34 $\pm$ 0.05	& 0.40 $\pm$ 0.04  	& 0.36 $\pm$ 0.04 	& 0.35 $\pm$ 0.07 	& 0.57 $\pm$ 0.18 \\
		& 3	& 0.29 $\pm$ 0.06	& 0.34 $\pm$ 0.05 	& 0.41 $\pm$ 0.04 	& 0.36 $\pm$ 0.04	& 0.36 $\pm$ 0.06	& 0.47 $\pm$ 0.14 \\
\hline
Ne/Fe		& 1 	& 1.34 $\pm$ 0.17	& 1.34 $\pm$ 0.15	& 1.83 $\pm$ 0.14  	& 1.54 $\pm$ 0.12  	& 1.00 $\pm$ 0.17 	& 1.4 $\pm$ 0.5 \\
		& 3	& 0.60 $\pm$ 0.15	& 0.73 $\pm$ 0.13	& 0.98 $\pm$ 0.10 	& 0.98 $\pm$ 0.10	& 0.56 $\pm$ 0.15	& $<$ 0.3 \\
\hline
Mg/Fe		& 1	& $<$ 0.4		& 0.66 $\pm$ 0.17	& 0.40 $\pm$ 0.12 	& $<$ 0.4 		& $<$ 0.5	 	& $<$ 0.2\\
		& 3	& 0.5 $\pm$ 0.2		& 1.19 $\pm$ 0.17	& 1.09 $\pm$ 0.12	& 0.68 $\pm$ 0.11	& 0.42 $\pm$ 0.18	& 1.0 $\pm$ 0.5 \\
\hline
Scale $s$		& 1	& 1.5 $\pm$ 0.2		& 0.98 $\pm$ 0.10	& 1.01 $\pm$ 0.10  	& 0.90 $\pm$ 0.07 	& 0.77 $\pm$ 0.13 	& 4.5 $\pm$ 1.0\\
		& 3	& 1.30 $\pm$ 0.16	& 0.78 $\pm$ 0.10	& 0.69 $\pm$ 0.06 	& 0.62 $\pm$ 0.05 	& 0.58 $\pm$ 0.11	& 3.0 $\pm$ 0.7 \\
\hline
$\chi^2$ / d.o.f.& 1	& 852 / 701		& 977 / 701		& 1491 / 701	 	& 1337 / 701	 	& 932 / 701		& 923 / 701\\
		& 3	& 757 / 698		& 802 / 698		& 800 / 698 		& 853 / 698		& 851 / 698		& 839 / 698 \\
\hline
\end{tabular}
\label{tab:rgs_sp}
\end{center}
\end{table*}

Because of the high statistics of the RGS spectrum shown in Fig.~\ref{fig:rgs_wdem}, we are also able to extract smaller strips in the cross-dispersion
direction and make a radial profile up to 2$\arcmin$ from the center. The results of these fits are shown in Table~\ref{tab:rgs_sp}. It confirms the
radial temperature profile in the core found with EPIC, although the temperature determined with RGS is slightly higher. The reason for this
discrepancy is probably the difference in extraction regions. The RGS spectra also contain a small contribution of photons originating outside the core
radius due to the spatial extent of the source. Again, like in the full field-of-view case, the mean temperatures $kT_{\mathrm{mean}}$ are slightly
higher than the single-temperature fits, although the values are consistent within the error-bars. The value for $\alpha$ is rather well constrained in
the region between $-2\arcmin$ and 0.5$\arcmin$ and shows a slight discontinuity around 0$\arcmin$, but the statistics do not allow to draw a
conclusion. From the $\chi^2$ values it is clear that a DEM model ({\it wdem}) fits the RGS spectra better. 

From the RGS abundance profiles in the cross-dispersion direction, which are listed in Table~\ref{tab:rgs_sp}, we confirm the spatial
distribution of oxygen and iron indicated by the line widths in the full RGS spectrum. The O/Fe ratio is consistent with a value between 0.3--0.4 
over the whole profile. The neon and magnesium abundance tend to peak in the center slightly more than iron. 
Because the individual spectra in the RGS spatial profile have lower statistics than the total RGS spectrum, we use one scale parameter $s$ for all measured lines.
Therefore, the fitted scale parameter, which accounts for the broadening of the lines because of the spatial extend of the source, is the average of all line
widths given in units of the width of cluster emission.
These widths increase naturally toward the outer part of the cluster, because the spatial profile of the cluster also flattens.

\section{Temperature and iron abundance maps}
\label{maps}

\begin{figure*}
\begin{minipage}{0.5\textwidth}
\includegraphics[width=0.85\textwidth,clip=t,angle=0.]{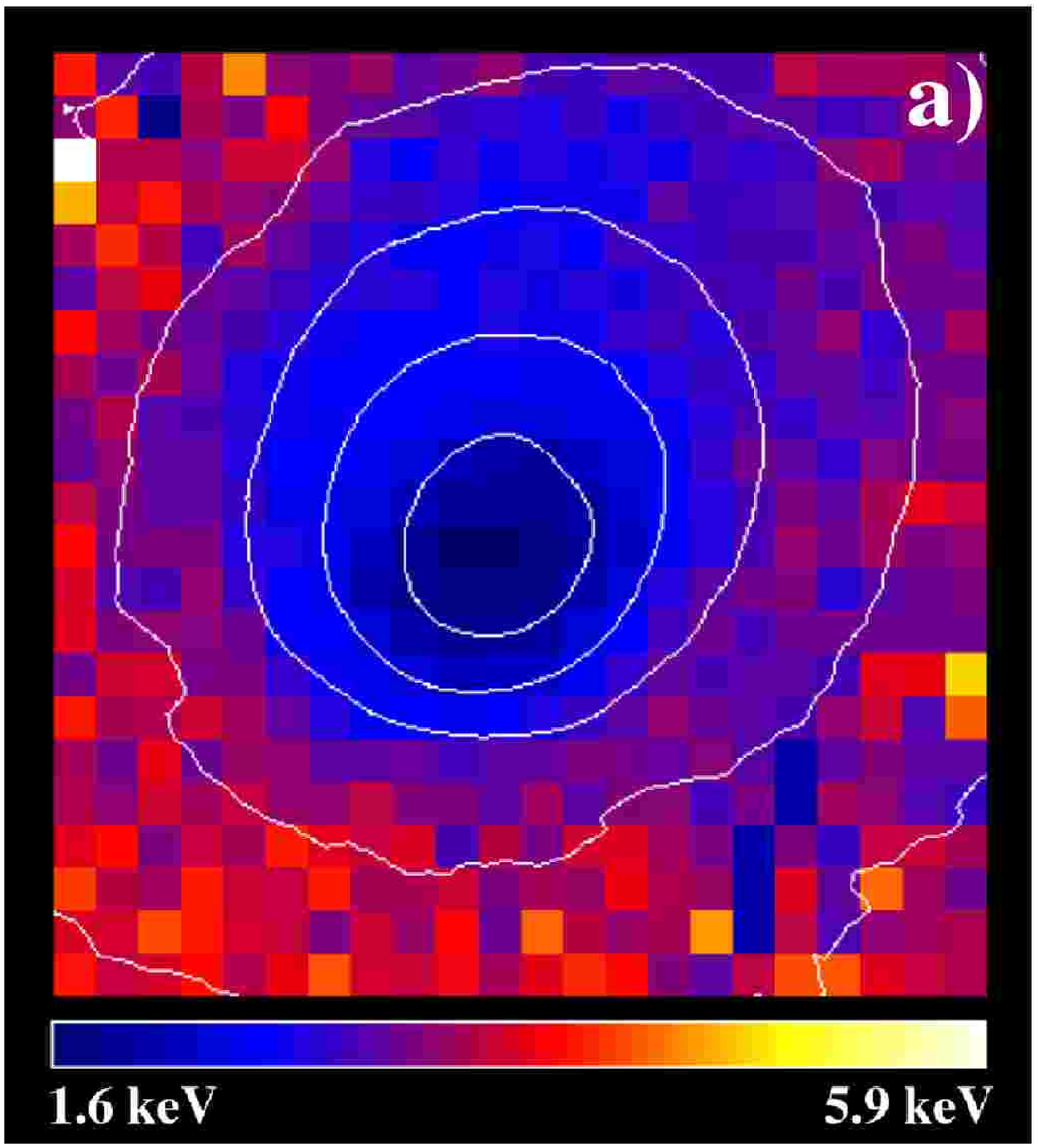}
\end{minipage}
\begin{minipage}{0.5\textwidth}
\includegraphics[width=0.85\textwidth,clip=t,angle=0.]{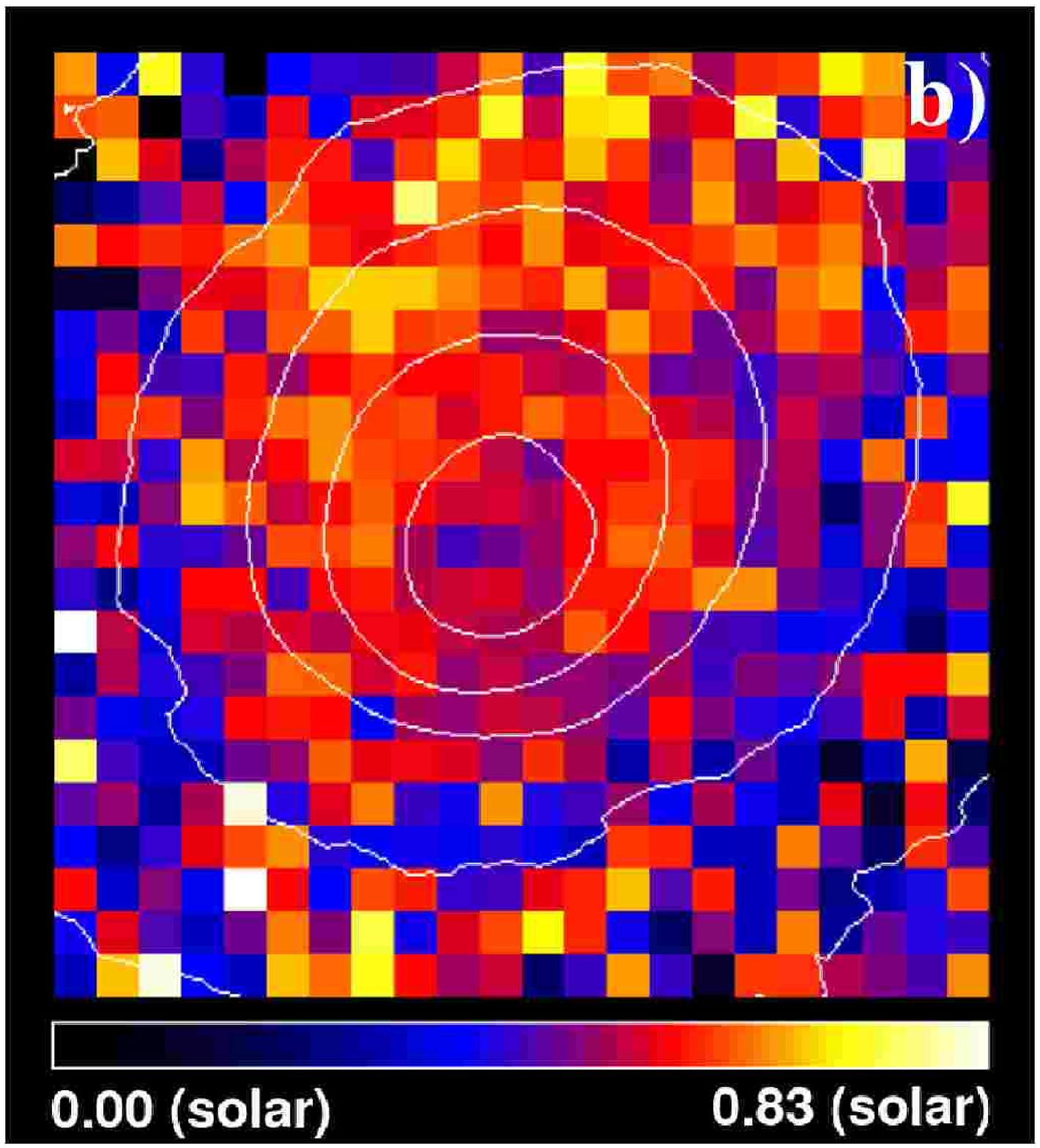}
\end{minipage}\\
\vspace{2mm}

\begin{minipage}{0.5\textwidth}
\includegraphics[width=0.85\textwidth,clip=t,angle=0.]{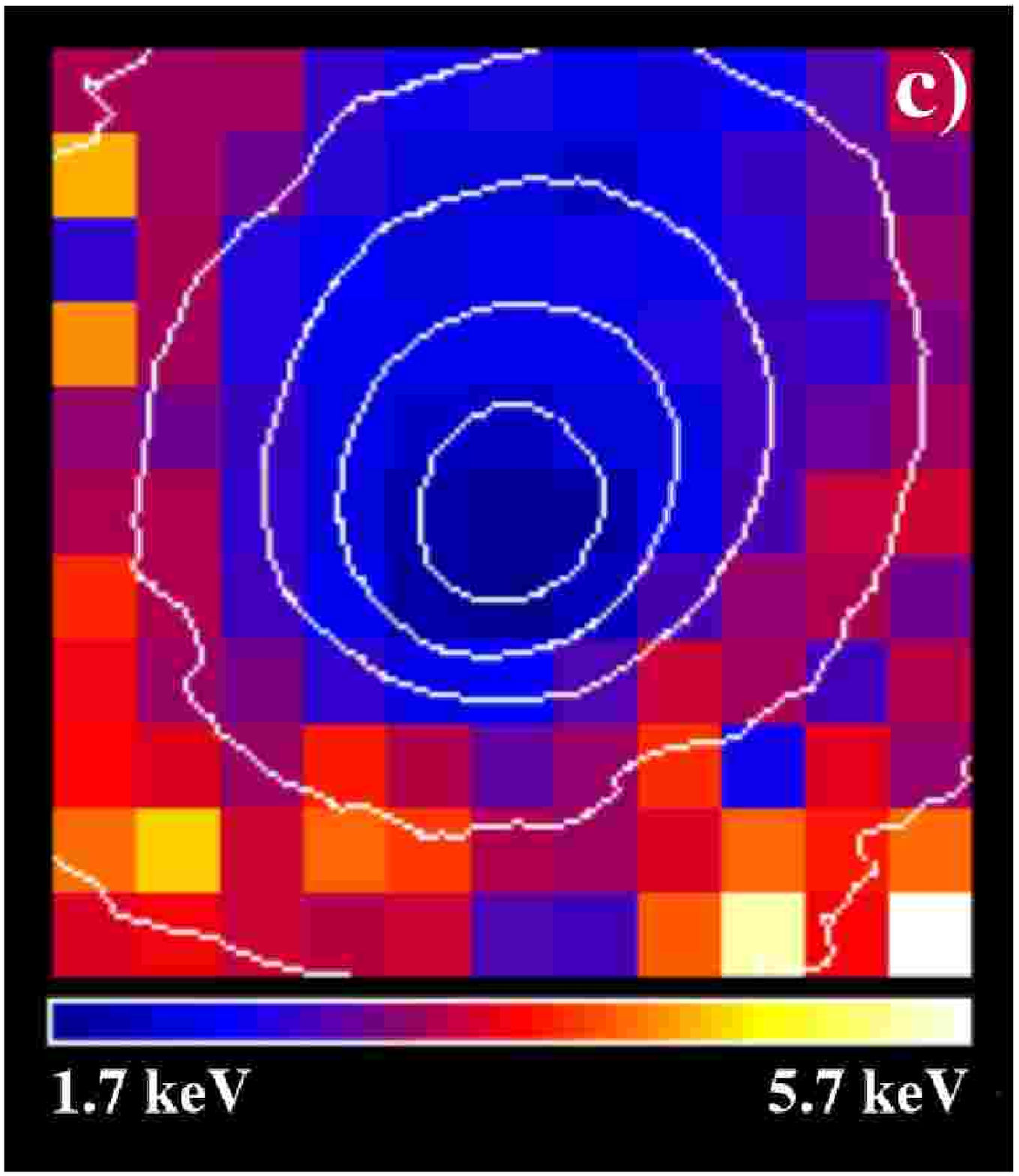}
\end{minipage}
\begin{minipage}{0.5\textwidth}
\includegraphics[width=0.85\textwidth,clip=t,angle=0.]{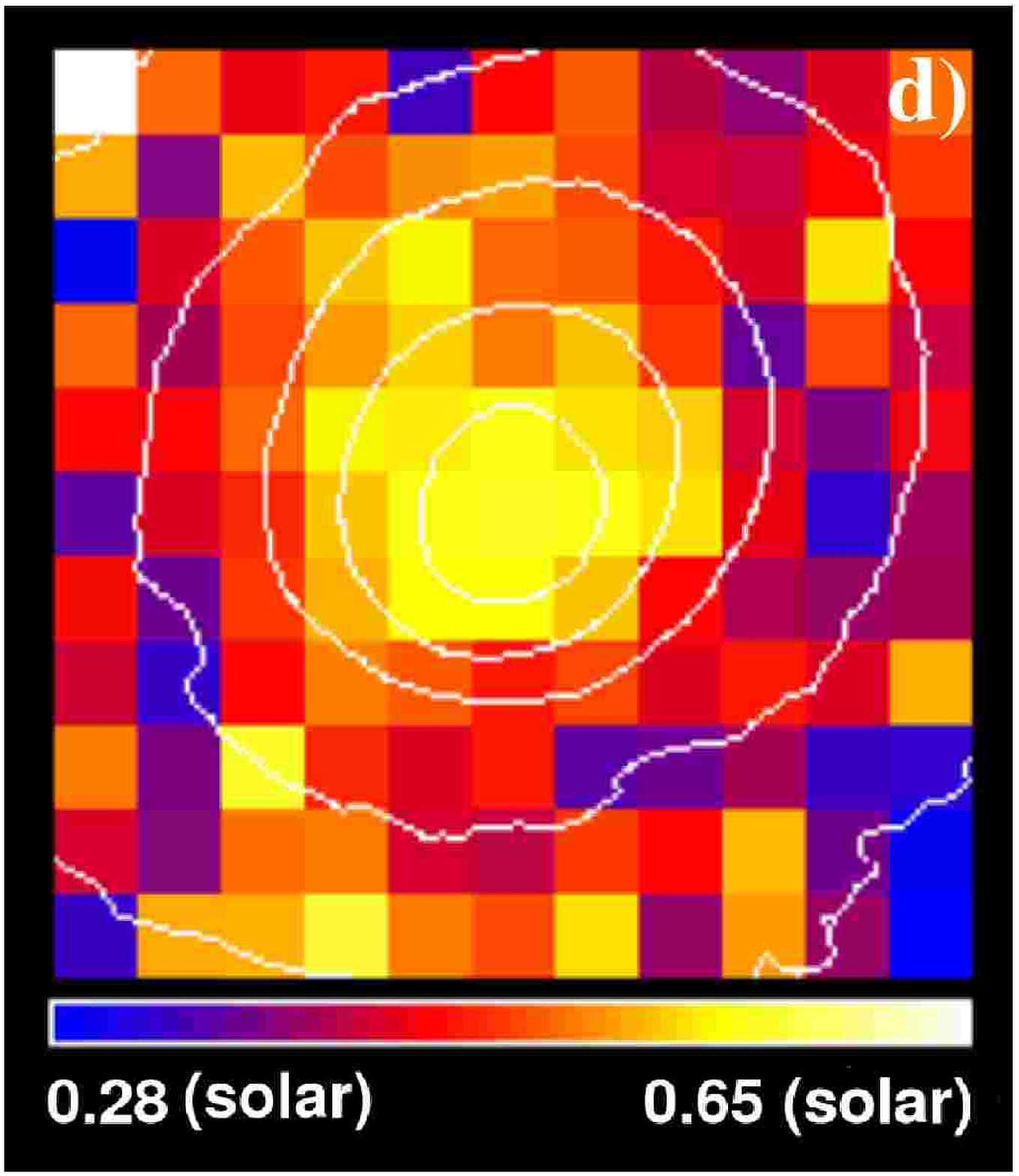}
\end{minipage}
\caption{Temperature and iron abundance maps of the inner $5.5\arcmin \times5.5\arcmin$ region of the cluster. Panel {\it{a}} and {\it{b}}: temperature and iron
abundance map with each $15\arcsec\times15\arcsec$ pixel fitted with a single-temperature model. Panel {\it{c}} and {\it{d}}: mean temperature and iron
abundance map with each $30\arcsec\times 30\arcsec$ pixel fitted with a {\it{wdem}} model. Note that North is up, West is to the right. All maps have overplotted the
same X-ray isophots.} 
\label{fig:maps}
\end{figure*}

\begin{figure}
\sidecaption
\includegraphics[width=\columnwidth]{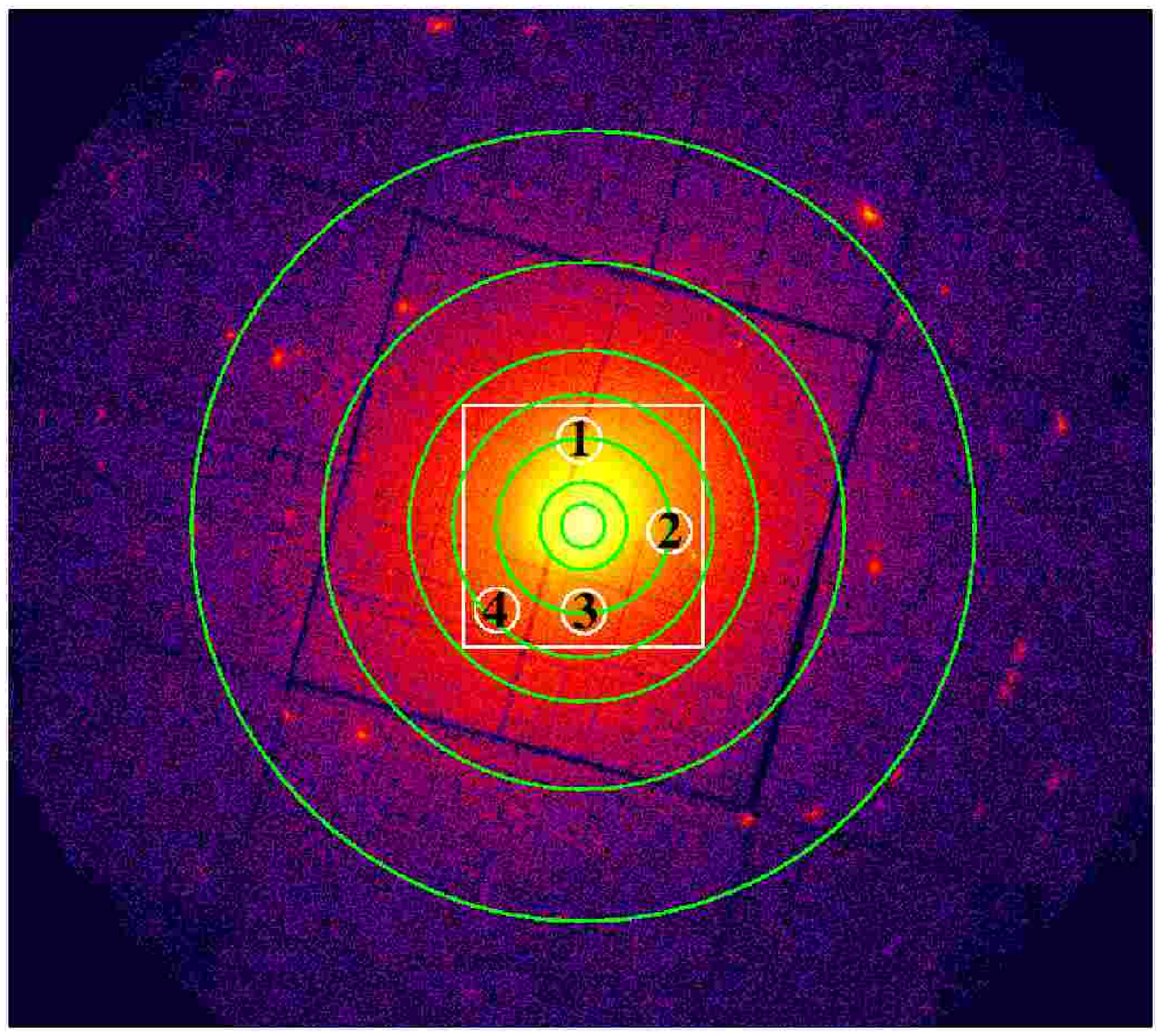}
\caption{MOS1 + MOS2 image of 2A~0335+096 with overplotted extraction regions. The concentric circles correspond to our extraction annuli used to investigate the radial
profiles, the square shows the $5.5\arcmin\times5.5\arcmin$ region used to extract the temperature and iron abundance maps and the small
circles correspond to extraction regions 1, 2, 3 and 4 used to verify that the temperature and abundances vary with position angle.}
\label{regions}
\end{figure}

In order to investigate the spatial variations of the temperature and metallicity we extract temperature and iron abundance maps of the $5.5\arcmin\times5.5\arcmin$ region
centered on the core of the cluster. 

First, we determine the temperature and iron abundance on a grid with bin-size of $30\arcsec\times30\arcsec$. 
For every bin we compute a redistribution and an ancillary file. We fit the spectrum of each bin individually by the multi-temperature {\it{wdem}} model.  The spectral fits
are done with a hydrogen column density fixed to the global value $2.5\times10^{21}$ cm$^{-2}$ (see Appendix \ref{infback}). The redshift is fixed to the spectroscopically
determined value of the central cD galaxy ($z=0.0349$). The abundances of all elements except iron in our 
model are fixed to 0.3 times the solar value (the exact value of these abundances does not influence the temperature and iron abundance determinations), while the iron
abundance is left as a free parameter. We fix the parameter $\alpha$ to 0.58, which is the value determined for the core of the cluster (see Table \ref{3models}). The free
parameters are the maximum temperature $T_{\mathrm{max}}$, the iron abundance and the normalization of the {\it{wdem}} component.  

The temperature map, which shows the mean temperature $T_{\mathrm{mean}}$ of the {\it{wdem}} model, reveals an elongated temperature structure in the South-Southeast
North-Northwest direction and a sharp temperature change over the brightness edge south of the core. The iron abundance map shows a highly centrally peaked 
iron abundance with an extension to the North (see the panel \emph{c} and \emph{d} in Fig. \ref{fig:maps}).    

High count rates in the $5.5\arcmin\times5.5\arcmin$ central region allow us to reduce the size of the bins to $15\arcsec\times15\arcsec$ (which is roughly the FWHM of the
point spread function of the pn detector) and extract a higher resolution temperature and iron abundance map fitting a single temperature model to each bin (see panel
{\it{a}} and {\it{b}} in Fig. \ref{fig:maps}). The map shows a sharp temperature change on the southern side of the core and a large extension of the cool temperature gas to
North-Northwest. In contrast to the {\it{wdem}} model fit, the iron abundance map extracted using a single-temperature model shows a central drop of iron abundance, with an
abundance peak in a "high metallicity ring" around the X-ray core with an extension to the North. We verified that the central drop of iron abundance is evident also in a
temperature map with $30\arcsec\times30\arcsec$ bins fitted with a single temperature model. The central abundance drop is rather the result of the oversimplified
model for the temperature structure of the core of the cluster, than a real feature (see subsection \ref{modelcomp}).  

To verify the azimuthal differences in the temperature and the metallicity we extract spectra from 4 circular regions with a radius of $0.5\arcmin$ and fit them with the
{\it{wdem}} model. The extraction regions are shown in Fig.~\ref{regions}. The regions 1, 2 and 3 are centered at the same distance from the core ($2\arcmin$) but at
different position angles (North, West, South). Region 4 is at a distance of $2.75\arcmin$ from the core at South-Southeast. Apart from the {\it{wdem}} normalization,
$T_{\mathrm{max}}$ and the iron abundance we leave free and fit the parameter $\alpha$, $N_{\mathrm{H}}$ and the silicon and sulfur abundance. The best-fit parameter
values are given in Table~\ref{mapregions}. We confirm the temperature and iron abundance asymmetry seen in the temperature maps.  

\begin{table}
\begin{center}
\caption[]{Parameter values obtained by fitting regions 1, 2, 3 and 4 indicated in Fig. \ref{regions} with the \emph{wdem} model. Emission measures ($Y = \int n_{\mathrm{e}}
n_{\mathrm{H}} dV$) are given in $10^{66}$ cm$^{-3}$, temperatures are given in keV, $N_{\mathrm{H}}$ in $10^{21}$ cm$^{-2}$ and abundances are given with respect to solar. 
\label{mapregions}}
\begin{tabular}{l|cccc}
\hline
\hline
Par.  & 1 (N)  & 2 (W) & 3 (S) & 4 (SSE) \\
\hline
Y & 4.55 & 2.63 & 1.84 & 1.16 \\
$N_{\mathrm{H}}$ & $2.54\pm 0.05$ & $2.32\pm0.1$ & $2.49\pm0.09$ & $2.38\pm0.17$\\
$kT_{\mathrm{max}}$ & $3.18\pm0.10$ & $4.15\pm0.20$ & $4.2\pm0.25$ & $4.9\pm0.4$\\
$kT_{\mathrm{mean}}$ & $2.69\pm0.13$ & $3.22\pm0.20$ & $3.25\pm0.24$ & $3.5\pm0.4$ \\
$\alpha$ & $0.22\pm0.06$ & $0.41\pm0.10$ & $0.41\pm0.11$ & $0.6\pm0.3$\\
Si         & $0.60\pm0.06$ & $0.70\pm0.10$ & $0.59\pm0.11$ & $0.50\pm0.24$\\   
S          & $0.46\pm0.08$ & $0.43\pm0.14$ & $0.17\pm0.14$ & $0.31\pm0.19$\\ 
Fe         & $0.51\pm0.02$ & $0.39\pm0.03$ & $0.36\pm0.03$ & $0.37\pm0.04$ \\
\hline
$\chi^{2}$ / d.o.f. & 605/509 & 512/509 & 551/509 & 598/509 \\
\hline
\end{tabular}
\noindent
\end{center}
\end{table}

\section{Properties of the core on smaller scales
\label{smallscale}}

\begin{figure}
\sidecaption
\includegraphics[width=\columnwidth]{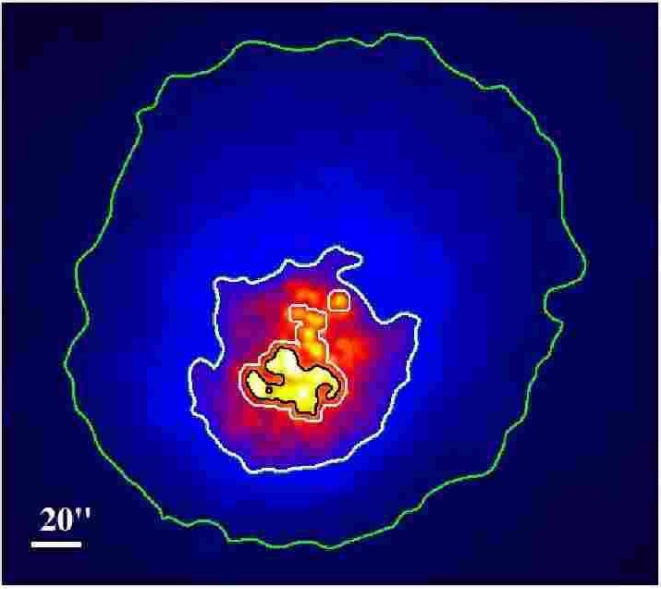}
\caption{Smoothed {\it{Chandra}} image of the core of 2A~0335+096 with overplotted areas used to extract spectra
from the data obtained during our deep XMM-Newton observation. The spectra of {\it{blobs/filaments}}, {\it{ambient gas}} and the {\it{envelope}} were
extracted respectively from the area indicated by black, white and gray contours.}
\label{centralim}
\end{figure}

The high statistics of the data obtained during our deep XMM-Newton observation allow us to better study the thermal properties of the blobs/filaments found in the core of
2A~0335+096 with Chandra \citep{Mazzotta2003}.
We use a smoothed Chandra ACIS image in the energy band of $0.5-1.5$ keV to locate the blobs/filaments. We create three extraction masks: one for
the region of bright {\it{blobs/filaments}}; one for the bright region around the blobs, but excluding the blobs ({\it{the ambient gas}}); and one for the bright
core but excluding the complex structure ({\it{the envelope}}); see Fig. \ref{centralim}. We use
these masks to extract spectra from our EPIC-MOS data. The Full Width Half Maximum (FWHM) of the on-axis Point Spread Function (PSF) of MOS1
and MOS2 at 1.5 keV is $4.3\arcsec$ and $4.4\arcsec$ respectively, while the FWHM of the on-axis PSF of EPIC-pn at the same energy is $12.5\arcsec$. Since the large PSF of EPIC-pn would
cause a strong contamination of the spectra extracted from the region of blobs/filaments, we analyze only the data obtained by MOS. 

\begin{figure}
\includegraphics[width=0.73\columnwidth,angle=270.]{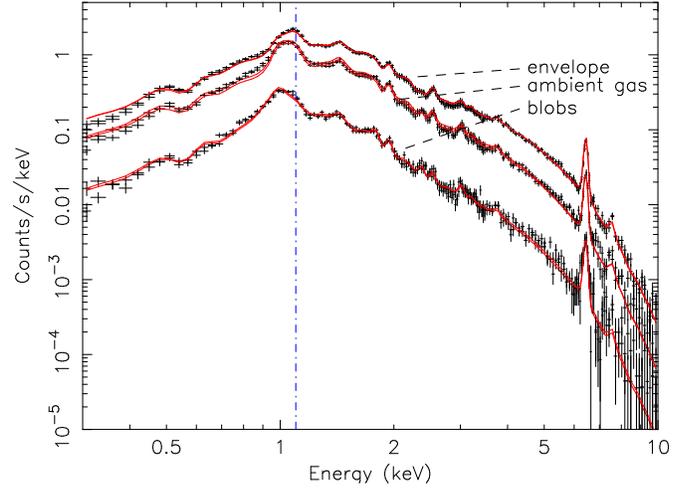}
\caption{Spectra of the {\it{blobs/filaments}}, the {\it{ambient gas}} around these features and of the {\it{envelope}} of gas
surrounding the inner core. The vertical line helps to see that there is more Fe-L emission at low energies in the spectrum of the blobs.}
\label{blobspectr}
\end{figure}

We fit the three spectra simultaneously. The envelope is fitted by a thermal model. The ambient inter-blob gas is fitted by two thermal models, while all
the parameters of the second thermal model are coupled to the model with which we fit the envelope. The region of blobs/filaments is fitted with three thermal models, where
the parameters of the second and third model are coupled respectively to the model we use to fit the ambient gas, and the envelope. 
This way we de facto deproject the
temperature of the blobs. The only free parameters in the fit are the normalizations of the thermal components, the temperatures, and the iron abundances. The values of the
other abundances are fixed to the values determined by fitting the central region with \emph{wdem}. The redshift is fixed to the redshift of the central cD
galaxy. The best fit parameters are shown in Table~\ref{blobtable} and the fitted spectra are shown in Fig. \ref{blobspectr}. The fit shows that the deprojected temperature
of the blobs/filaments is $1.14\pm0.01$ keV, while the temperature of the ambient gas is $1.86\pm0.01$ keV. The iron abundance of the blobs is coupled to the iron abundance
of the ambient gas, while the iron abundance of the envelope is fitted separately and is found to be somewhat lower. Coupling the iron abundance of the blobs with that of
the ambient gas is necessary to avoid anti-correlation between the two parameters and it has no effect on the determined temperature difference between the two components.

\citet{Mazzotta2003} assumed that the blobs are ellipsoids and selected eight regions. We assume the regions and the luminosities given by \citet{Mazzotta2003} and with our
new temperature we calculate with a MEKAL model the density of the blobs. We find a value of $\sim8.5\times10^{-2}$ cm$^{-3}$. From our best fit emission measure of the
thermal model of the ambient gas we calculate the density of the gas and find a value of $\sim5.5\times10^{-2}$ cm$^{-3}$. These density estimates and our new values of the
temperatures of these two components therefore indicate, that the cold blobs/filaments and the ambient gas are in thermal pressure equilibrium
($nkT\simeq0.1$~keV~cm$^{-3}$).  

\begin{table}
\begin{center}
\caption[]{Temperature and iron abundance of the {\it{blobs/filaments}}, the {\it{ambient gas}} around these features and of the {\it{envelope}} of gas
surrounding the inner core. The superscript $a$ indicates that the parameters are coupled.}
\label{blobtable}
\begin{tabular}{llll}
\hline
\hline
& blo./fil. & amb. gas & envelope \\
\hline
$kT$   & $1.14\pm0.01$ & $1.86\pm0.01$ & $2.79\pm0.01$ \\
Fe (solar)     & $0.63^{a}\pm0.02$ & $0.63^{a}\pm0.02$ & $0.61\pm0.01$ \\         
\hline
\end{tabular}
\end{center}
\end{table}

\section{Abundances and enrichment by supernova types Ia/II and Population III stars}

\begin{figure}
\sidecaption
\includegraphics[width=\columnwidth,clip=t,angle=0.]{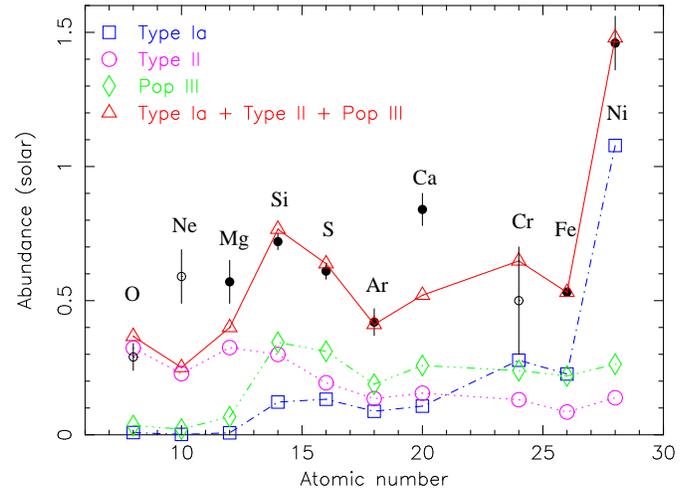}
\caption{The elemental abundances derived from a {\it{wdem}} fit of the inner $3\arcmin$ of 2A~0335+096, fitted by a linear combination of the yields of
SN~Ia, SN~II and Pop~III, for the SN~Ia model W7 and Pop~III model with a core mass of 130~M$_{\odot}$. The black empty circles indicate elements not used in the fit.} 
\label{fig:supernova}
\end{figure}

\begin{figure}
\sidecaption
\includegraphics[width=\columnwidth,clip=t,angle=0]{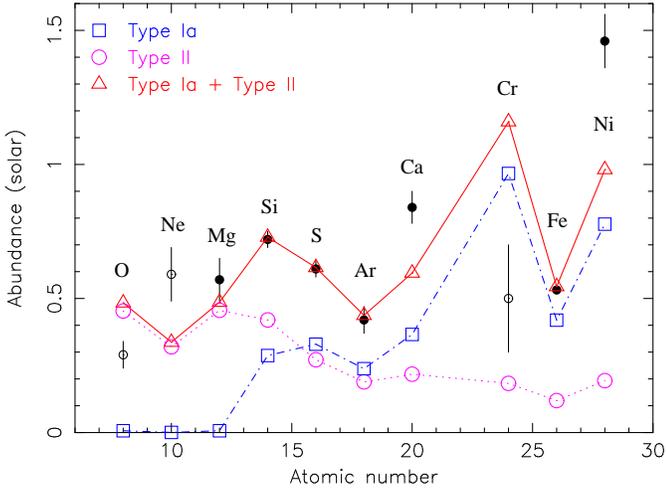}
\caption{The elemental abundances derived from a {\it{wdem}} fit of the inner $3\arcmin$ of 2A~0335+096, fitted by a linear combination of the yields of
SN~Ia and SN~II. For SN~Ia we use the WDD2 model. The black empty circles indicate elements not used in the fit.} 
\label{fig:supernova2}
\end{figure}

\begin{table*}
\caption{Relative numbers of SN~Ia, SN~II and Pop~III contributing to the enrichment of the intra-cluster medium. We show the results of fits of linear combinations of the
yields of SN~Ia, SN~II and Pop~III stars, with two SN~Ia yield models, to the abundances of 7 elements shown in Table~\ref{global} determined
for the core and the outer region of 2A~0335+096. For the core we show the results of fits with two different Pop~III star models and the results of fits without Pop~III stars.}
\begin{center}
\begin{tabular}{l|l|cc|cc|cc|cc}
\hline\hline
Model   & Type	& \multicolumn{6}{c}{$0\arcmin-3\arcmin$}               & \multicolumn{2}{c}{$3\arcmin-9\arcmin$} \\
\hline

        &        &$M_{\mathrm{popIII}}=130$~M$_{\odot}$& $\chi^2$ / d.o.f.&$M_{\mathrm{popIII}}=65$ M$_{\odot}$& $\chi^2$ / d.o.f. &no Pop III& $\chi^2$ / d.o.f. &no Pop III & $\chi^2$ / d.o.f. \\ 

\hline
W7	& SN~Ia	& $0.24\pm0.04$ &		   & $0.18\pm0.01$ &         & $0.20\pm0.01$ &         & $0.39\pm0.04$& \\
	& SN~II	& $0.76\pm0.04$ & 36 / 4	   & $0.85\pm0.01$ & 60 / 4  & $0.80\pm0.01$ & 70 / 5  & $0.61\pm0.04$ & 44/5 \\
	& Pop~III& $(4\pm1)\times10^{-3}$ &	   & $(-2.7\pm0.4)\times10^{-2}$    &    & $-$           &	       & $-$ & \\ 
\hline
WDD2	& SN~Ia	& $0.25\pm0.04$ &		   &$0.27\pm0.03$&           &$0.26\pm0.01$ &        & $0.37\pm0.04$ & \\
	& SN~II 	& $0.75\pm0.04$ & 35 / 4   &$0.72\pm0.03$& 43 / 4    &$0.74\pm0.01$ & 43 / 5  & $0.63\pm0.04$ & 8/5\\
	& Pop~III& $(-2\pm1)\times10^{-3}$ &       &$(6\pm18)\times10^{-3}$ &       &  $-$            &      & $-$ & \\   
\hline
\end{tabular}
\end{center}
\label{tab:supernovafit}
\end{table*}

We investigate the relative contribution of the number of supernova types Ia/II to the total enrichment of the
intra-cluster medium (ICM). We also investigate the possibility of putting constraints on the contribution by Population~III stars.
We try to determine the relative numbers of different supernovae
\begin{equation}
\frac{N_{\mathrm{Sx}}}{N_{\mathrm{Ia}}+N_{\mathrm{II}}+N_{\mathrm{popIII}}},
\end{equation}
where $N_{\mathrm{Sx}}$ is either the number of Type~Ia supernovae (SNe~Ia), Type~II supernovae (SNe~II) or Popupation~III stars (Pop~III)
contributing to the enrichment of ICM. 
Since the progenitors of Type II supernovae and Type Ib and Ic (SNe~Ib/Ic) are massive stars, we associate the heavy-element
yields from SNe~II with those from SNe~Ib/Ic. For nucleosynthesis products of SN~II we adopt an average yield of stars on a mass 
range from 10~M$_\odot$ to 50~M$_\odot$ calculated by \citet{tsujimoto1995} assuming a Salpeter initial mass function. For nucleosynthesis
products of SN~Ia we adopt values calculated by  \citet{iwamoto1999} and investigate two different models (see Table~\ref{tab:supernovafit}).
The W7 model is calculated using a slow deflagration model, while the WDD2 model is calculated using a delayed-detonation model. Note that
delayed-detonation models are currently favored over deflagration models by the supernova community, whereas the WDD2 is favored by
\citet{iwamoto1999}. The intergalactic medium might also have been significantly enriched by Pop~III stars which exploded as pair instability
supernovae. We investigate two models for heavy element yields of Pop~III stars calculated by \citet{heger2002}, with a core mass of 130
M$_{\odot}$ and 65 M$_{\odot}$. 

We assume that the observed total number of atoms $N_{i}$ of the element $i$ (determined from its abundance and the emission measure distribution) is a linear
combination of the number of atoms $Y_{i}$ produced per individual supernova Type~Ia (Y$_{i,\mathrm{Ia}}$), Type~II (Y$_{i,\mathrm{II}}$) and Population~III star
(Y$_{i,\mathrm{III}}$): 
\begin{equation}
N_i = a\mathrm{Y}_{i,\mathrm{Ia}} + b\mathrm{Y}_{i,\mathrm{II}} + c\mathrm{Y}_{i,\mathrm{III}},
\end{equation}
where $a$, $b$ and $c$ are multiplication factors representing the total number of each supernovae that went off in the cluster and enriched the ICM.

First we fit the abundances obtained for the cooling-core region (the inner $3\arcmin$) and
determine the best-fit values for $a$, $b$ and $c$, which we use to estimate the relative numbers of supernovae. We fit the abundance values
obtained for the inner $3\arcmin$ region shown in Table~\ref{global}. 
The best fit values of the relative contributions are shown in Table~\ref{tab:supernovafit} and Fig.~\ref{fig:supernova}. 
In the Fig.~\ref{fig:supernova} we also show our best measurement of the chromium abundance and the measured oxygen and neon abundances
determined by the RGS and normalized to the iron abundance determined by EPIC (the absolute abundances are not well determined by RGS). Since
the low significance makes the value of the chromium abundance uncertain and the abundance values of the oxygen and neon may be affected by
systematic uncertainties arising from using different extraction regions for the RGS and EPIC, we do not include them in the fitting. We see
that all models are consistent with a scenario where the relative number of Type~Ia supernovae contributing to the enrichment of the
intra-cluster medium is $20-30$\%, while the relative number of Type~II supernovae is $70-80$\%. 

The 130~M$_{\odot}$ Pop~III star model fits show, that their relative number contributing to the enrichment is about two orders of
magnitude lower than the number of SN~Ia/II. However, they produce according to the models about two orders of magnitude more mass per star
than SN~Ia/II so their contribution to the enrichment of ICM could still be significant. However, the relative numbers determined for Pop~III
are model dependent and the high values of the $\chi^2$s do not allow us to confirm their contribution.
The 65~M$_{\odot}$ Pop~III star model fits give unphysical values for their relative numbers. 
In Fig.~\ref{fig:supernova2} we show our best-fit model of the observed abundances as a linear combination of only SN~Ia and SN~II. This
model gives us the same relative number of SN~Ia as the models containing Pop~III stars. Also, using the WDD2 SN~Ia model, the $\chi^2$ does
not change with respect to models containing Pop~III stars. 
The fit results show that Pop~III stars are not necessary to explain the observed
abundance patterns, although their contribution to the enrichment of the ICM cannot be excluded. We conclude that this analysis does not
allow us to put any constrains on the enrichment by Pop~III stars.

In Fig.~\ref{fig:supernova} and ~\ref{fig:supernova2} we see that the measured abundance of calcium is significantly higher than that
predicted by the models. The abundance of nickel is well fitted only using the W7 model of Type~Ia supernovae with Pop~III stars. 
Fitting the abundances as a linear combination of SN~Ia and SN~II, the W7 model predicts more and the WDD2 model predicts less nickel than observed. 
Mainly the discrepancies between the model and the measured abundances of calcium and nickel are responsible for the large $\chi^2$s. 
The WDD2 model always predicts significantly more chromium than the upper limit derived from our data-point. The best measured abundances 
of oxygen and neon are also not well reconstructed by the models. 
This may be either due to problems in the supernova models or due to the fact that they were
determined from a different extraction region with a different instrument.   
We note that the involved uncertainties and complications when interpreting the observed abundances \citep[see][]{matteucci2005} do not allow us to
differentiate between the two SN~Ia models.

Since our fits of the core show that we cannot put constrains on the contribution of Pop~III stars we fit the ``outskirts''
($3\arcmin-9\arcmin$) with models containing only SN~Ia and SN~II. 
The fits show a higher contribution of SN~Ia in the outer region than in the cooling core region, however the large $\chi^2$ values
and the involved systematic uncertainties do not allow us to make a firm statement.

\section{Discussion}
\label{dis}

We have used three different thermal models in our analysis of 2A~0335+096 and find that the multi-temperature {\it{wdem}} model fits the data best. We note that a good
description of the temperature structure is very important to obtain correct values for elemental abundances. 
Fitting the EPIC and RGS data we determine the radial temperature and abundance profiles for several elements. The elemental abundances have a peak in the core of the cluster
and show a gradient across the whole cluster. The abundance structure of the cluster is
consistent with a scenario where the relative number of SN~Ia is $20-30$\% of the total number of supernovae enriching the ICM. Contrary to the findings of
\citet{baumgartner2005}, the Population~III stars are not necessary to explain the observed abundance patterns. However, their contribution to the enrichment of the ICM
cannot be excluded.
Extracting temperature and iron abundance maps we find an asymmetry across the cluster. We find that the deprojected temperature of the blobs/filaments in the core of the
cluster is lower than the temperature of the ambient gas and the blobs appear to be in thermal pressure equilibrium with the ambient gas. 
Below, we discuss the implications of these results.

\subsection{Intrinsic temperature structure}

Our results show a presence of plasma emitting at a range of different temperatures within every extraction bin. In the core of the cluster, this may be caused in part by
projection effects and by the strong temperature gradient.  
The multi-temperature structure we find around the central bin may also be caused in part by the large scale temperature asymmetry of the
cluster, so in one extraction annulus we extract the spectrum of both the colder and of the warmer gas from North and South of the cluster center respectively. 
However, the non-isothermality detected by fitting the deprojected spectra cannot be caused only by the width of
our extraction annuli in combination with the temperature asymmetry. The temperature gradients across our annuli and the temperature differences between North and South are
not large enough for that. The results can be interpreted as intrinsic multi-temperature structure at each position, which is in agreement with the results of
\citet{kaastra2004}.  Furthermore, these results show that fitting the spectrum of the intra-cluster gas with a single-temperature thermal model is in general not sufficient
and introduces systematic errors especially in the determination of elemental abundances. 

\subsubsection{Spatial temperature distribution}

The temperature distribution has a strong gradient in the core. In the inner $3\arcmin$ the deprojected temperature drops by a factor of $\sim 1.8$ toward the center. In the
outer part ($3\arcmin-9\arcmin$) the radial temperature distribution is flat. We do not see a temperature drop toward the outskirts.
We find an asymmetry in the temperature distribution and iron abundance distribution. The temperature maps show the asymmetry in temperature distribution in the direction of
the elongated surface brightness morphology, roughly in the South-Southeast North-Northwest direction. The temperature drops
faster toward the core from the South than from the North. The temperature change is the strongest over the cold front South of the core identified by \citet{Mazzotta2003}
in Chandra data. The cold fronts are interpreted as boundaries of a dense, cooler gas moving through the hotter, more rarefied ambient gas \citep{Markevitch2000}. From the
pressure jump across the cold front \citet{Mazzotta2003} conclude that the dense central gas core moves from North to South with a Mach number of $M\simeq0.75\pm0.2$. The
elongated, comet-like shape of the cold central core shown in our temperature maps supports this picture. The elongated shape of the cold gas core may hint that we see a
surviving core of a merged subcluster moving through and being stripped by a less dense surrounding gas. The possible merger scenario is discussed in subsection
\ref{mergersec}.

\subsubsection{X-ray blobs/filaments}

\citet{Mazzotta2003} found that the deprojected temperature of the blobs is consistent with that of the less
dense ambient gas, so these gas phases do not appear to be in thermal pressure equilibrium. Therefore they propose a
possibility of a significant unseen non-thermal pressure component in the inter-blob gas, possibly arising from the activity of a central
active galactic nucleus (AGN). They also discuss two models to explain the origin of the blobs: hydrodynamic instabilities caused by the
motion of the cool core and ``bubbling'' of the core caused by multiple outbursts of the central AGN. 

However, the temperature measurements of \citet{Mazzotta2003} have large uncertainties due to the low
statistics of the Chandra data. The superior spectral quality of our XMM-Newton data allows us to measure the temperature of the blobs much
more accurately. In contrast to the previous findings we find that the deprojected temperature of the region where they identified the
blobs/filaments is colder by a factor of 1.6 than the temperature of the ambient gas and the blobs appear to be in thermal pressure
equilibrium with the ambient gas. In the light of this new result a non-thermal pressure component in the inter-blob gas is not necessary to explain the
observed blobs. 

\subsection{Abundance distribution
\label{abundances}}

The high statistics of the data allow us to determine the abundances of the most abundant metals. For the first time we put constraints on the abundance of chromium,
and upper limits on abundances of titanium, manganese and cobalt. The abundance distribution has a peak in the core of the cluster. The radial profiles show an abundance
gradient toward the core across the whole analyzed region of the cluster. The abundance gradient is not restricted to the cooling-flow region. 
We observe centrally peaked abundance distributions for both SN~Ia products (iron, nickel) and elements contributed mainly by SN~II (magnesium, silicon). 
Contrary to our results from 2A~0335+096, cooling core clusters (e.g. see the sample of \citet{tamura2004}) in general show a centraly peaked distribution of SN~Ia
products with respect to SN~II products. The different abundance distribution in 2A~0335+096 might point toward a different enrichment history.
In S\'ersic 159-03 the widths of the RGS lines also indicate that the iron abundance is much more centrally peaked than oxygen \citep{deplaa2005b}. 
Contrary to S\'ersic 159-03, the RGS data in 2A~0335+096 indicate that the spatial distributions of oxygen and iron might be very similar. However, we have to be careful with
interpreting the RGS results, because the strong temperature drop in the center of this cluster might cause that the RGS spectrum is dominated by the line emission from the cool
core and the line profiles of iron and oxygen follow the temperature structure more than the abundance distribution of the elements.

We estimate that the total mass of iron within the radius of 126 kpc is $\sim2.1\times10^9$~M$_{\odot}$. If all iron originates from SN~Ia, we need $\sim2.8\times10^9$ SN~Ia
explosions to explain the observed amount of iron. Assuming $1.1\times10^{10}$ years and 50 galaxies contributing iron via galactic winds and gas stripping we get an average
SN~Ia rate of one explosion per 200 years per galaxy. 

We show that the determined abundance values are sensitive to the applied temperature model. Especially single-temperature thermal models can introduce a bias in the
abundance determination, resulting in a detection of off-center abundance peaks reported for example by \citet{sanders2002} and \citet{johnstone2002}.

The abundance maps show an asymmetry in the iron abundance distribution, with a higher abundance to the North of the cluster core. The abundance asymmetry further
highlights and supports the picture of great dynamical complexity. 

\subsection{Possible merger scenario}
\label{mergersec}

We observe an asymmetry of the temperature and iron abundance distribution across the cluster, along the surface brightness elongation, which points toward
the possibility of an ongoing merger in 2A~0335+096.
\citet{Mazzotta2003} note that the orientation of the cold front and the asymmetry of the temperature distribution suggest that the cool core is moving along a
projected direction that goes from North to South. However, it is not known whether the motion is due to a merger, or due to gas sloshing induced by an off-axis merger or by
a flyby of another cluster.  
\citet{Mazzotta2003} also reason that if the cool core is indeed a merging subcluster, then in projection the position of the subcluster is close to
the cluster center. However, the impact parameter cannot be measured hence the subcluster may be passing through the cluster at any distance along the line of sight.
Our detection of an asymmetry in the abundance distribution across the cluster, with a higher metallicity North of the core with respect to the South, supports the merger
scenario, in which the subcluster had a higher metallicity. The most important enrichment processes, like ram pressure stripping of cluster galaxies
and galactic winds, are not expected to produce the observed asymmetry in the spatial distribution of the iron abundance.  In the
dense cores of clusters of galaxies the internal kinematics of the ICM is expected to quickly smooth out any existing inhomogeneities in the
abundance distribution.

\begin{figure}
\includegraphics[width=\columnwidth]{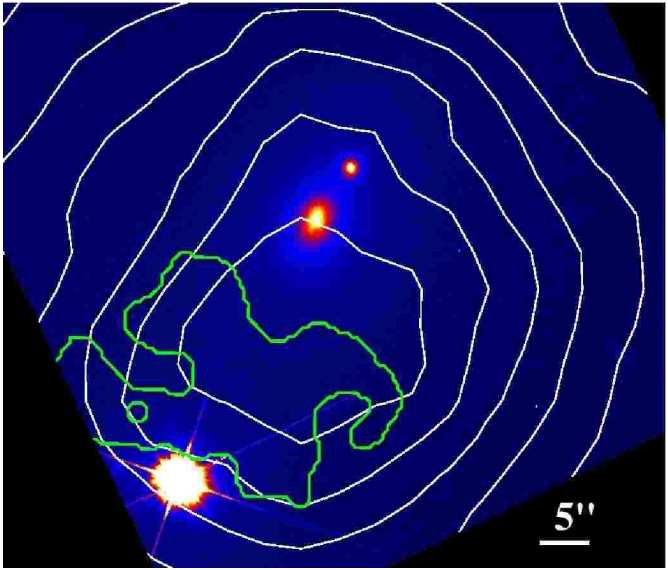}
\caption{An image of the core of the cluster obtained by 
  the Hubble Space Telescope with overplotted X-ray isophots. North is up, West is right. The X-ray isophots are off-set from the central
  cD~galaxy. The gray contour indicates the location of the cold blobs/filaments detected in X-rays.}
\label{hubble}
\end{figure}

In Fig.~\ref{hubble} we show an archival \emph{Hubble Space Telescope} (HST) image of the core of the cluster with overplotted EPIC-MOS X-ray isophots. The HST image
shows the central cD galaxy and a companion galaxy to the North-West (the bright source in the lower left is a foreground star). We see, that the
X-ray brightness peak is offset to the South from the central cD~galaxy, which also shows that the core is not in a relaxed state. This was already shown by
\citet{Mazzotta2003} who found that the
centroid positions of the best fit double $\beta$-model to the cluster surface brightness do not coincide with the position of the central cD~galaxy, but are offset to the
South-East and North-West. It is interesting to note, that the alignment of the central cD~galaxy and the alignment of the central galaxy with the projected companion
nucleus are following the same direction as the possible merger axis inferred from the X-rays. \citet{dubinski1998} has shown that the long axis
of a brightest cluster galaxy always aligns with the primordial filament along which the cluster collapsed and with the long axis of the cluster galaxy distribution. If
2A~0335+096 formed along a filament aligned as the long axis of the central cD galaxy, we might be seeing the traces of the latest phase of the
formation of the cluster, in which a subcluster is falling in along the same primordial filament.

Although there is growing evidence for the merger scenario, with the available data we still cannot rule out an alternative scenario of gas sloshing. 
Radial velocity measurements of the individual galaxies in 2A~0335+096 are necessary to provide the decisive evidence.

\subsection{Relative enrichment by supernova types Ia/II and Population III stars}

All our fits show that the contribution of SN~Ia is $20-30$\% of the total number of supernovae enriching the ICM.
\citet{tsujimoto1995} found that the ratio of the total numbers (integrated over time) of supernovae Type Ia to Type II that best reproduces the Galactic abundances is
$N_{\mathrm{Ia}}/N_{\mathrm{II}}=0.15$, while the ratio determined for the Magellanic Clouds is $0.2-0.3$, which correspond to relative numbers of
$N_{\mathrm{Ia}}/N_{\mathrm{Ia+II}}=0.13$ and $0.17-0.23$ in the 
Galaxy and in the Magellanic clouds respectively. Supernova types Ia/II produce heavy-elements on different time scales during the chemical evolution of galaxies, with SN~II
causing the enrichment usually in the early phases of galactic evolution and SNe~Ia on a much longer time-scale, in the later phases of galactic evolution. The current
relative frequencies of SNe~Ia and II also depend on types of galaxies. 
According to the Lick Observatory Supernova Search (LOSS) about 42\% of the supernovae observed in galaxies within a redshift of $z=0.03$ are SN~Ia, while
SNe~II are rare in early-type galaxies and SNe~Ia occur in all types of galaxies \citep{bergh2005}. 
The relative contribution to the enrichment of ICM by SN~Ia in the cluster of galaxies 2A~0335+096 is between the Galactic value and the value
of the current relative frequencies determined by LOSS \citep{bergh2005}.
The value of $N_{\mathrm{Ia}}/N_{\mathrm{Ia+II}}=20-30$\% is consistent with a picture of an early enrichment by SNe~II material from
star-burst galaxies and later contribution by Type~Ia supernovae material. 

However, the scenario proposed by \citet{tamura2001b} and \citet{finoguenov2001}, where an early and well mixed SN~II enrichment
is followed by SN~Ia enrichment in the central galaxy, produces over the Hubble time a centrally peaked distribution of SN~Ia products
relative to SN~II products. As discussed in \ref{abundances}, we observe a centrally peaked abundance distribution of both SN~Ia and SN~II
products.  Moreover, contrary to the expectations, comparing the relative enrichment in the inner and outer regions of the cluster the Mg/Fe,
Si/Fe and S/Fe ratios point toward a higher contribution of SN~Ia in the outer parts compared to the cool core. Such a difference in the 
enrichment history between the core and the outer parts would strongly support the scenario according to which the cool core is the
surviving ICM of a merged subcluster (see \ref{mergersec}). However, the involved uncertainties in the determination of the relative enrichment
are high.  

As discussed by \citet{matteucci2005} the interpretation of the observed abundance ratios is not easy. The observed abundance ratios strongly
depend on the star-formation history in cluster galaxies. 
Part of the supernova products might still be locked in the stars or in the potential wells of cluster galaxies.
Therefore, the relative number of supernovae contributing to the enrichment of the ICM, which we infer from the abundance ratios, might be
somewhat different from the relative number of supernovae which exploded in the cluster over its life-time.

The results also show that it is not necessary to include the Pop~III stars in our models to explain the observed abundance
patterns. Our magnesium, silicon, sulfur, argon and iron abundances can be reconstructed with a linear combination of yields of SN~Ia and II
for all tested SN~Ia yield models. In the cluster 2A~0335+096 we cannot confirm the results of \citet{baumgartner2005} claiming, that
especially in the hotter clusters, Pop~III stars are necessary to explain the observed silicon, sulfur, calcium and argon abundances. However, to test the
results of \citet{baumgartner2005} an analysis of a large cluster sample including hotter clusters will be necessary.

All fitted models give a significantly lower calcium abundance than we observe. This points toward uncertainties in the models of SN~II
yields, e.g. different explosion energies in the models can result in a difference of an order of magnitude in calcium production
\citep{woosley1995}.

\section{Conclusions}
We have analyzed spatially resolved and high resolution spectra of the cluster of galaxies 2A~0335+096 obtained during a deep XMM-Newton
observation. We found that: 
\begin{itemize}
\item
We unambiguously detect multi-temperature structure in the core of 2A~0335+096: for the determination of elemental abundances, a
multi-temperature model is mandatory. The {\it{wdem}} model is a good description of the temperature structure of the gas in 2A~0335+096. 
\item
The blobs/filaments found in the core of 2A~0335+096 are significantly colder than the ambient gas and they appear to be in pressure
equilibrium with their environment.
\item
We detect a strong central peak in the abundance distributions of both SN~Ia and SN~II products.
\item
The relative number of SN~Ia contributing to the enrichment of the intra-cluster medium in the central 130~kpc of 2A~0335+096 is $\sim25$\%,
while the relative number of SN~II is $\sim75$\%. Comparison of the observed abundances to the supernova yields does not allow us
to put any constrains on the contribution of Pop~III stars to the enrichment of the ICM. We observe significantly higher calcium abundance
than predicted by supernova models.  
\item
The detected asymmetry in the temperature and iron abundance distribution further supports the merger scenario, in which the subcluster had
a higher metallicity.

\end{itemize}

\begin{acknowledgements}
This work is based on observations obtained with XMM-Newton, an ESA science mission with instruments
and contributions directly funded by ESA member states and the USA (NASA). The Netherlands Institute
for Space Research (SRON) is supported financially by NWO, the Netherlands Organization for Scientific 
Research.
\end{acknowledgements}

\bibliographystyle{aa}
\bibliography{3868}

\begin{thebibliography}{60}
\expandafter\ifx\csname natexlab\endcsname\relax\def\natexlab#1{#1}\fi

\bibitem[{{Allende Prieto} {et~al.}(2001){Allende Prieto}, {Lambert}, \&
  {Asplund}}]{allende2001}
{Allende Prieto}, C., {Lambert}, D.~L., \& {Asplund}, M. 2001, \apjl, 556, L63

\bibitem[{{Anders} \& {Grevesse}(1989)}]{anders1989}
{Anders}, E. \& {Grevesse}, N. 1989, \gca, 53, 197

\bibitem[{{Baumgartner} {et~al.}(2005){Baumgartner}, {Loewenstein}, {Horner},
  \& {Mushotzky}}]{baumgartner2005}
{Baumgartner}, W.~H., {Loewenstein}, M., {Horner}, D.~J., \& {Mushotzky}, R.~F.
  2005, \apj, 620, 680

\bibitem[{{Bonamente} {et~al.}(2005){Bonamente}, {Lieu}, {Mittaz}, {Kaastra},
  \& {Nevalainen}}]{bonamente2005}
{Bonamente}, M., {Lieu}, R., {Mittaz}, J.~P.~D., {Kaastra}, J.~S., \&
  {Nevalainen}, J. 2005, \apj, 629, 192

\bibitem[{{Buote}(2000)}]{Buote2000}
{Buote}, D.~A. 2000, \mnras, 311, 176

\bibitem[{{Cooke} {et~al.}(1978){Cooke}, {Ricketts}, {Maccacaro}, {Pye},
  {Elvis}, {Watson}, {Griffiths}, {Pounds}, {McHardy}, {Maccagni}, {Seward},
  {Page}, \& {Turner}}]{Cooke1978}
{Cooke}, B.~A., {Ricketts}, M.~J., {Maccacaro}, T., {et~al.} 1978, \mnras, 182,
  489

\bibitem[{{Davis}(2001)}]{davis2001}
{Davis}, J.~E. 2001, \apj, 548, 1010

\bibitem[{{De Grandi} \& {Molendi}(2001)}]{degrandi2001}
{De Grandi}, S. \& {Molendi}, S. 2001, \apj, 551, 153

\bibitem[{{De Grandi} \& {Molendi}(2002)}]{degrandi2002}
{De Grandi}, S. \& {Molendi}, S. 2002, \apj, 567, 163

\bibitem[{{De Luca} \& {Molendi}(2004)}]{deluca2004}
{De Luca}, A. \& {Molendi}, S. 2004, \aap, 419, 837

\bibitem[{{de Plaa} {et~al.}(2005){de Plaa}, {Kaastra}, {M\'endez}, {Tamura},
  {Bleeker}, {Peterson}, {Paerels}, {Bonamente}, \& {Lieu}}]{deplaa2005a}
{de Plaa}, J., {Kaastra}, J.~S., {M\'endez}, M., {et~al.} 2005, ASR, in press.

\bibitem[{{de Plaa} {et~al.}(2004){de Plaa}, {Kaastra}, {Tamura},
  {Pointecouteau}, {Mendez}, \& {Peterson}}]{deplaa2004}
{de Plaa}, J., {Kaastra}, J.~S., {Tamura}, T., {et~al.} 2004, \aap, 423, 49

\bibitem[{{de Plaa} {et~al.}(2006){de Plaa}, {Werner}, {Bykov}, {Kaastra},
  {M\'endez}, {Vink}, {Bleeker}, {Bonamente}, \& {Peterson}}]{deplaa2005b}
{de Plaa}, J., {Werner}, N., {Bykov}, A., {et~al.} 2006, \aap, submitted

\bibitem[{{den Herder} {et~al.}(2001){den Herder}, {Brinkman}, {Kahn},
  {Branduardi-Raymont}, {Thomsen}, {Aarts}, {Audard}, {Bixler}, {den Boggende},
  {Cottam}, {Decker}, {Dubbeldam}, {Erd}, {Goulooze}, {G{\" u}del},
  {Guttridge}, {Hailey}, {Janabi}, {Kaastra}, {de Korte}, {van Leeuwen},
  {Mauche}, {McCalden}, {Mewe}, {Naber}, {Paerels}, {Peterson}, {Rasmussen},
  {Rees}, {Sakelliou}, {Sako}, {Spodek}, {Stern}, {Tamura}, {Tandy}, {de
  Vries}, {Welch}, \& {Zehnder}}]{herder2001}
{den Herder}, J.~W., {Brinkman}, A.~C., {Kahn}, S.~M., {et~al.} 2001, \aap,
  365, L7

\bibitem[{{Dickey} \& {Lockman}(1990)}]{dickey1990}
{Dickey}, J.~M. \& {Lockman}, F.~J. 1990, \araa, 28, 215

\bibitem[{{Dubinski}(1998)}]{dubinski1998}
{Dubinski}, J. 1998, \apj, 502, 141

\bibitem[{{Edge}(2001)}]{edge2001}
{Edge}, A.~C. 2001, \mnras, 328, 762

\bibitem[{{Ettori} {et~al.}(2002){Ettori}, {De Grandi}, \&
  {Molendi}}]{Ettori2002}
{Ettori}, S., {De Grandi}, S., \& {Molendi}, S. 2002, \aap, 391, 841

\bibitem[{{Fabian}(1994)}]{fabian1994}
{Fabian}, A.~C. 1994, \araa, 32, 277

\bibitem[{{Finoguenov} {et~al.}(2001){Finoguenov}, {Arnaud}, \&
  {David}}]{finoguenov2001}
{Finoguenov}, A., {Arnaud}, M., \& {David}, L.~P. 2001, \apj, 555, 191

\bibitem[{{Grevesse} \& {Sauval}(1998)}]{Grevesse1998}
{Grevesse}, N. \& {Sauval}, A.~J. 1998, Space Science Reviews, 85, 161

\bibitem[{{Heger} \& {Woosley}(2002)}]{heger2002}
{Heger}, A. \& {Woosley}, S.~E. 2002, \apj, 567, 532

\bibitem[{{Irwin} \& {Sarazin}(1995)}]{Irwin1995}
{Irwin}, J.~A. \& {Sarazin}, C.~L. 1995, \apj, 455, 497

\bibitem[{{Iwamoto} {et~al.}(1999){Iwamoto}, {Brachwitz}, {Nomoto},
  {Kishimoto}, {Umeda}, {Hix}, \& {Thielemann}}]{iwamoto1999}
{Iwamoto}, K., {Brachwitz}, F., {Nomoto}, K., {et~al.} 1999, \apjs, 125, 439

\bibitem[{{Jansen} {et~al.}(2001){Jansen}, {Lumb}, {Altieri}, {Clavel}, {Ehle},
  {Erd}, {Gabriel}, {Guainazzi}, {Gondoin}, {Much}, {Munoz}, {Santos},
  {Schartel}, {Texier}, \& {Vacanti}}]{jansen2001}
{Jansen}, F., {Lumb}, D., {Altieri}, B., {et~al.} 2001, \aap, 365, L1

\bibitem[{{Johnstone} {et~al.}(2002){Johnstone}, {Allen}, {Fabian}, \&
  {Sanders}}]{johnstone2002}
{Johnstone}, R.~M., {Allen}, S.~W., {Fabian}, A.~C., \& {Sanders}, J.~S. 2002,
  \mnras, 336, 299

\bibitem[{{Kaastra} {et~al.}(2001){Kaastra}, {Ferrigno}, {Tamura}, {Paerels},
  {Peterson}, \& {Mittaz}}]{kaastra2001}
{Kaastra}, J.~S., {Ferrigno}, C., {Tamura}, T., {et~al.} 2001, \aap, 365, L99

\bibitem[{{Kaastra} {et~al.}(2003){Kaastra}, {Lieu}, {Tamura}, {Paerels}, \&
  {den Herder}}]{kaastra2003a}
{Kaastra}, J.~S., {Lieu}, R., {Tamura}, T., {Paerels}, F.~B.~S., \& {den
  Herder}, J.~W. 2003, \aap, 397, 445

\bibitem[{{Kaastra} {et~al.}(1996){Kaastra}, {Mewe}, \&
  {Nieuwenhuijzen}}]{kaastra1996}
{Kaastra}, J.~S., {Mewe}, R., \& {Nieuwenhuijzen}, H. 1996, in UV and X-ray
  Spectroscopy of Astrophysical and Laboratory Plasmas p.411, K. Yamashita and
  T. Watanabe. Tokyo : Universal Academy Press

\bibitem[{{Kaastra} {et~al.}(2004){Kaastra}, {Tamura}, {Peterson}, {Bleeker},
  {Ferrigno}, {Kahn}, {Paerels}, {Piffaretti}, {Branduardi-Raymont}, \&
  {B\"ohringer}}]{kaastra2004}
{Kaastra}, J.~S., {Tamura}, T., {Peterson}, J.~R., {et~al.} 2004, \aap, 413,
  415

\bibitem[{{Kikuchi} {et~al.}(1999){Kikuchi}, {Furusho}, {Ezawa}, {Yamasaki},
  {Ohashi}, {Fukazawa}, \& {Ikebe}}]{Kikuchi1999}
{Kikuchi}, K., {Furusho}, T., {Ezawa}, H., {et~al.} 1999, \pasj, 51, 301

\bibitem[{{Kuntz} \& {Snowden}(2000)}]{kuntz2000}
{Kuntz}, K.~D. \& {Snowden}, S.~L. 2000, \apj, 543, 195

\bibitem[{{Lodders}(2003)}]{lodders2003}
{Lodders}, K. 2003, \apj, 591, 1220

\bibitem[{{Lumb} {et~al.}(2002){Lumb}, {Warwick}, {Page}, \& {De
  Luca}}]{lumb2002}
{Lumb}, D.~H., {Warwick}, R.~S., {Page}, M., \& {De Luca}, A. 2002, \aap, 389,
  93

\bibitem[{{Markevitch} {et~al.}(2000){Markevitch}, {Ponman}, {Nulsen}, {Bautz},
  {Burke}, {David}, {Davis}, {Donnelly}, {Forman}, {Jones}, {Kaastra},
  {Kellogg}, {Kim}, {Kolodziejczak}, {Mazzotta}, {Pagliaro}, {Patel}, {Van
  Speybroeck}, {Vikhlinin}, {Vrtilek}, {Wise}, \& {Zhao}}]{Markevitch2000}
{Markevitch}, M., {Ponman}, T.~J., {Nulsen}, P.~E.~J., {et~al.} 2000, \apj,
  541, 542

\bibitem[{{Matteucci} \& {Chiappini}(2005)}]{matteucci2005}
{Matteucci}, F. \& {Chiappini}, C. 2005, Publications of the Astronomical
  Society of Australia, 22, 49

\bibitem[{{Mazzotta} {et~al.}(2003){Mazzotta}, {Edge}, \&
  {Markevitch}}]{Mazzotta2003}
{Mazzotta}, P., {Edge}, A.~C., \& {Markevitch}, M. 2003, \apj, 596, 190

\bibitem[{{Molendi} \& {Gastaldello}(2001)}]{Molendi2001}
{Molendi}, S. \& {Gastaldello}, F. 2001, \aap, 375, L14

\bibitem[{{Moretti} {et~al.}(2003){Moretti}, {Campana}, {Lazzati}, \&
  {Tagliaferri}}]{moretti}
{Moretti}, A., {Campana}, S., {Lazzati}, D., \& {Tagliaferri}, G. 2003, \apj,
  588, 696

\bibitem[{{Peterson} {et~al.}(2003){Peterson}, {Kahn}, {Paerels}, {Kaastra},
  {Tamura}, {Bleeker}, {Ferrigno}, \& {Jernigan}}]{peterson2003}
{Peterson}, J.~R., {Kahn}, S.~M., {Paerels}, F.~B.~S., {et~al.} 2003, \apj,
  590, 207

\bibitem[{{Peterson} {et~al.}(2001){Peterson}, {Paerels}, {Kaastra}, {Arnaud},
  {Reiprich}, {Fabian}, {Mushotzky}, {Jernigan}, \& {Sakelliou}}]{peterson2001}
{Peterson}, J.~R., {Paerels}, F.~B.~S., {Kaastra}, J.~S., {et~al.} 2001, \aap,
  365, L104

\bibitem[{{Read} \& {Ponman}(2003)}]{read2003}
{Read}, A.~M. \& {Ponman}, T.~J. 2003, \aap, 409, 395

\bibitem[{{Romanishin} \& {Hintzen}(1988)}]{Romanishin1988}
{Romanishin}, W. \& {Hintzen}, P. 1988, \apjl, 324, L17

\bibitem[{{Sanders} \& {Fabian}(2002)}]{sanders2002}
{Sanders}, J.~S. \& {Fabian}, A.~C. 2002, \mnras, 331, 273

\bibitem[{{Sarazin} {et~al.}(1995){Sarazin}, {Baum}, \& {O'Dea}}]{sarazin1995}
{Sarazin}, C.~L., {Baum}, S.~A., \& {O'Dea}, C.~P. 1995, \apj, 451, 125

\bibitem[{{Sarazin} {et~al.}(1992){Sarazin}, {O'Connell}, \&
  {McNamara}}]{Sarazin1992}
{Sarazin}, C.~L., {O'Connell}, R.~W., \& {McNamara}, B.~R. 1992, \apjl, 397,
  L31

\bibitem[{{Schwartz} {et~al.}(1980){Schwartz}, {Schwarz}, \&
  {Tucker}}]{Schwartz1980}
{Schwartz}, D.~A., {Schwarz}, J., \& {Tucker}, W. 1980, \apjl, 238, L59

\bibitem[{{Singh} {et~al.}(1986){Singh}, {Westergaard}, \&
  {Schnopper}}]{Singh1986}
{Singh}, K.~P., {Westergaard}, N.~J., \& {Schnopper}, H.~W. 1986, \apjl, 308,
  L51

\bibitem[{{Singh} {et~al.}(1988){Singh}, {Westergaard}, \&
  {Schnopper}}]{Singh1988}
{Singh}, K.~P., {Westergaard}, N.~J., \& {Schnopper}, H.~W. 1988, \apj, 331,
  672

\bibitem[{{Str{\" u}der} {et~al.}(2001){Str{\" u}der}, {Briel}, {Dennerl},
  {Hartmann}, {Kendziorra}, {Meidinger}, {Pfeffermann}, {Reppin}, {Aschenbach},
  {Bornemann}, {Br{\" a}uninger}, {Burkert}, {Elender}, {Freyberg}, {Haberl},
  {Hartner}, {Heuschmann}, {Hippmann}, {Kastelic}, {Kemmer}, {Kettenring},
  {Kink}, {Krause}, {M{\" u}ller}, {Oppitz}, {Pietsch}, {Popp}, {Predehl},
  {Read}, {Stephan}, {St{\" o}tter}, {Tr{\" u}mper}, {Holl}, {Kemmer},
  {Soltau}, {St{\" o}tter}, {Weber}, {Weichert}, {von Zanthier},
  {Carathanassis}, {Lutz}, {Richter}, {Solc}, {B{\" o}ttcher}, {Kuster},
  {Staubert}, {Abbey}, {Holland}, {Turner}, {Balasini}, {Bignami}, {La
  Palombara}, {Villa}, {Buttler}, {Gianini}, {Lain{\' e}}, {Lumb}, \&
  {Dhez}}]{struder2001}
{Str{\" u}der}, L., {Briel}, U., {Dennerl}, K., {et~al.} 2001, \aap, 365, L18

\bibitem[{{Tamura} {et~al.}(2001{\natexlab{a}}){Tamura}, {Bleeker}, {Kaastra},
  {Ferrigno}, \& {Molendi}}]{tamura2001b}
{Tamura}, T., {Bleeker}, J.~A.~M., {Kaastra}, J.~S., {Ferrigno}, C., \&
  {Molendi}, S. 2001{\natexlab{a}}, \aap, 379, 107

\bibitem[{{Tamura} {et~al.}(2004){Tamura}, {Kaastra}, {den Herder}, {Bleeker},
  \& {Peterson}}]{tamura2004}
{Tamura}, T., {Kaastra}, J.~S., {den Herder}, J.~W.~A., {Bleeker}, J.~A.~M., \&
  {Peterson}, J.~R. 2004, \aap, 420, 135

\bibitem[{{Tamura} {et~al.}(2001{\natexlab{b}}){Tamura}, {Kaastra}, {Peterson},
  {Paerels}, {Mittaz}, {Trudolyubov}, {Stewart}, {Fabian}, {Mushotzky}, {Lumb},
  \& {Ikebe}}]{tamura2001a}
{Tamura}, T., {Kaastra}, J.~S., {Peterson}, J.~R., {et~al.} 2001{\natexlab{b}},
  \aap, 365, L87

\bibitem[{{Tsujimoto} {et~al.}(1995){Tsujimoto}, {Nomoto}, {Yoshii},
  {Hashimoto}, {Yanagida}, \& {Thielemann}}]{tsujimoto1995}
{Tsujimoto}, T., {Nomoto}, K., {Yoshii}, Y., {et~al.} 1995, \mnras, 277, 945

\bibitem[{{Turner} {et~al.}(2001){Turner}, {Abbey}, {Arnaud}, {Balasini},
  {Barbera}, {Belsole}, {Bennie}, {Bernard}, {Bignami}, {Boer}, {Briel},
  {Butler}, {Cara}, {Chabaud}, {Cole}, {Collura}, {Conte}, {Cros}, {Denby},
  {Dhez}, {Di Coco}, {Dowson}, {Ferrando}, {Ghizzardi}, {Gianotti}, {Goodall},
  {Gretton}, {Griffiths}, {Hainaut}, {Hochedez}, {Holland}, {Jourdain},
  {Kendziorra}, {Lagostina}, {Laine}, {La Palombara}, {Lortholary}, {Lumb},
  {Marty}, {Molendi}, {Pigot}, {Poindron}, {Pounds}, {Reeves}, {Reppin},
  {Rothenflug}, {Salvetat}, {Sauvageot}, {Schmitt}, {Sembay}, {Short},
  {Spragg}, {Stephen}, {Str{\" u}der}, {Tiengo}, {Trifoglio}, {Tr{\" u}mper},
  {Vercellone}, {Vigroux}, {Villa}, {Ward}, {Whitehead}, \&
  {Zonca}}]{turner2001}
{Turner}, M.~J.~L., {Abbey}, A., {Arnaud}, M., {et~al.} 2001, \aap, 365, L27

\bibitem[{{van den Bergh} {et~al.}(2005){van den Bergh}, {Li}, \&
  {Filippenko}}]{bergh2005}
{van den Bergh}, S., {Li}, W., \& {Filippenko}, A.~V. 2005, \pasp, in press.

\bibitem[{{Weisskopf} {et~al.}(2004){Weisskopf}, {O'Dell}, {Paerels}, {Becker},
  {Tennant}, \& {Swartz}}]{weisskopf2004}
{Weisskopf}, M.~C., {O'Dell}, S.~L., {Paerels}, F., {et~al.} 2004, \apj, 601,
  1050

\bibitem[{{White} {et~al.}(1991){White}, {Fabian}, {Johnstone}, {Mushotzky}, \&
  {Arnaud}}]{White1991}
{White}, D.~A., {Fabian}, A.~C., {Johnstone}, R.~M., {Mushotzky}, R.~F., \&
  {Arnaud}, K.~A. 1991, \mnras, 252, 72

\bibitem[{{Woosley} \& {Weaver}(1995)}]{woosley1995}
{Woosley}, S.~E. \& {Weaver}, T.~A. 1995, \apjs, 101, 181

\bibitem[{{Zwicky} {et~al.}(1965){Zwicky}, {Karpowicz}, \&
  {Kowal}}]{Zwicky1965}
{Zwicky}, F., {Karpowicz}, M., \& {Kowal}, C.~T. 1965, in \emph{Catalog of
  Galaxies and Clusters of Galaxies} Vol. 5 (Pasadena: California Institute of
  Technology)

\end{thebibliography}

\appendix

\section{The influence of the background subtraction on radial profiles
\label{infback}}

\begin{figure}
\psfig{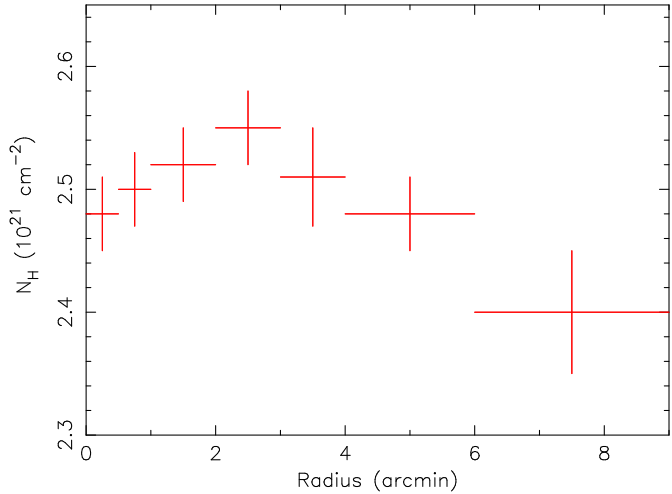}
\caption{The radial hydrogen column density profile.}
\label{NH}
\end{figure}

\begin{figure*}
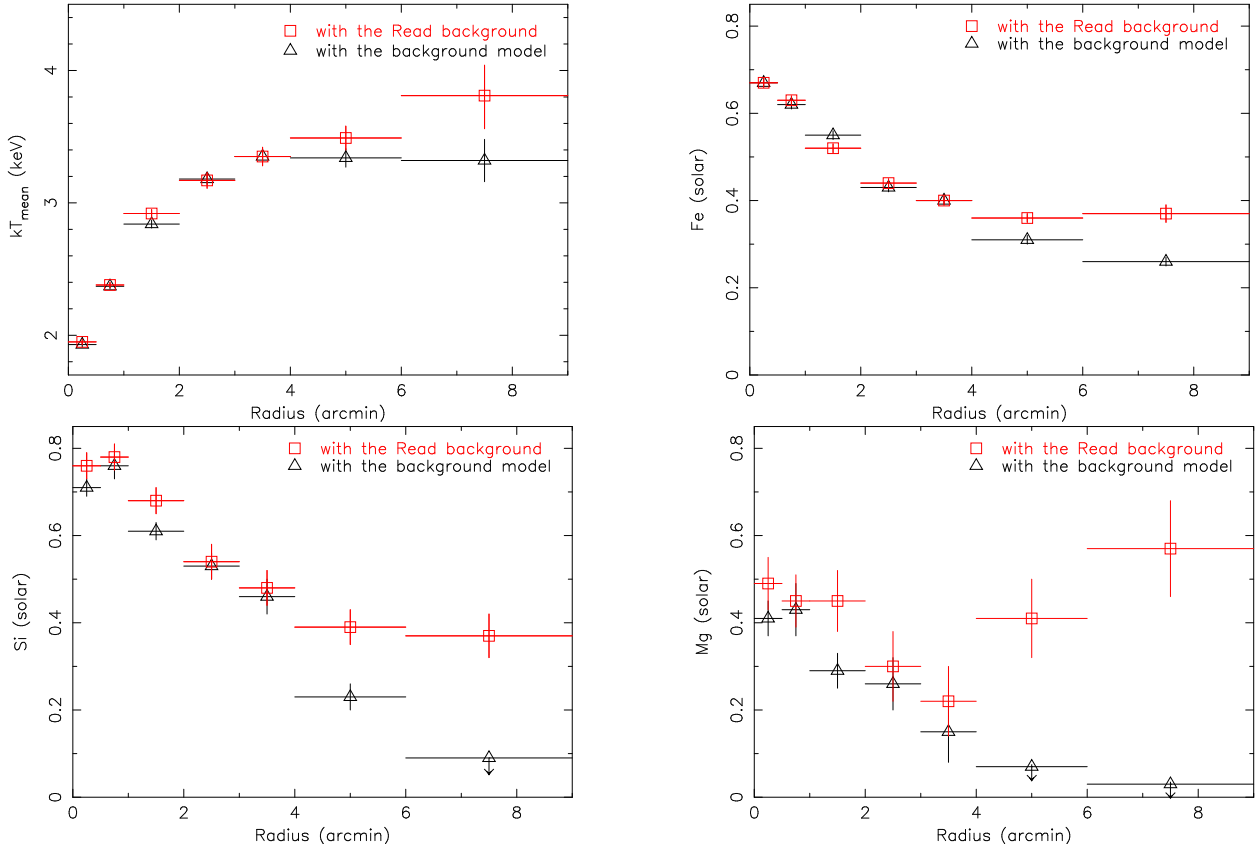

\begin{minipage}{0.5\textwidth}
\includegraphics[width=0.62\textwidth,clip=t,angle=270.]{3868fA2a.ps}
\end{minipage}
\begin{minipage}{0.5\textwidth}
\includegraphics[width=0.62\textwidth,clip=t,angle=270.]{3868fA2b.ps}
\end{minipage}\\
\vspace{2mm}
\begin{minipage}{0.5\textwidth}
\includegraphics[width=0.62\textwidth,clip=t,angle=270.]{3868fA2c.ps}
\end{minipage}
\begin{minipage}{0.5\textwidth}
\includegraphics[width=0.62\textwidth,clip=t,angle=270.]{3868fA2d.ps}
\end{minipage}
\caption{Comparison of fit results obtained by fitting the {\it{wdem}} model to EPIC spectra using the combined background event files of \citet{read2003}
and using our new background subtraction method.}
\label{fig:MvR}
\end{figure*}

\begin{table*}
\begin{center}
\caption[]{Fit results obtained by fitting the {\it{wdem}} model to EPIC spectra using ({\it{R}}) the combined background event lists of
\citet{read2003} and using ({\it{M}}) the new background subtraction method. Abundances are given  with respect to solar.}
\label{tab:MvR}
\begin{tabular}{l|c|ccccccc}
\hline
\hline
 Parameter & backg.  & $0-0.5\arcmin $ & $0.5-1.0\arcmin $  & $1.0-2.0\arcmin $ & $2.0-3.0\arcmin $ & $3.0-4.0\arcmin $ & $4.0-6.0\arcmin $ &  $6.0-9.0\arcmin $ \\
\hline
$N_{\mathrm{H}}$ ($10^{21} $cm$^{-2}$) & R & $2.48\pm0.03$ & $2.50\pm0.03$ & $2.52\pm0.03$ & $2.55\pm0.03$ & $2.51\pm0.04$ & $2.48\pm0.03$ & $2.40\pm0.05$\\
\hline
$kT_{\mathrm{max}}$ (keV)     & R & $2.66\pm0.02$ & $3.16\pm0.02$ & $3.98\pm0.04$ & $4.40\pm0.06$ & $4.73\pm0.07$ & $5.04\pm0.10$ & $6.28\pm0.23$ \\
$kT_{\mathrm{max}}$ (keV)     & M & $2.64\pm0.02$ & $3.14\pm0.02$ & $3.95\pm0.03$ & $4.39\pm0.04$ & $4.81\pm0.05$ & $4.97\pm0.05$ & $5.84\pm0.14$ \\
\hline
$kT_{\mathrm{mean}}$ (keV)    & R & $1.95\pm0.02$ & $2.38\pm0.02$ & $2.92\pm0.03$ & $3.17\pm0.06$ & $3.35\pm0.07$ & $3.49\pm0.09$ & $3.81\pm0.25$  \\
$kT_{\mathrm{mean}}$ (keV)    & M & $1.93\pm0.02$ & $2.37\pm0.02$ & $2.84\pm0.03$ & $3.18\pm0.04$ & $3.35\pm0.04$ & $3.34\pm0.07$ & $3.32\pm0.16$  \\
\hline
$\alpha$                      & R & $0.57\pm0.01$ & $ 0.49\pm0.02 $ & $0.57\pm0.02$ & $0.64\pm0.05$ & $0.71\pm0.06$ & $0.83\pm0.07$ & $2.4^{+0.8}_{-0.5}$ \\
$\alpha$                      & M & $0.58\pm0.01$ & $0.48\pm0.02$ & $0.65\pm0.02$ & $0.62\pm0.04$ & $0.79\pm0.04$ & $1.01^{+0.23}_{-0.02}$ & $2.37^{+13}_{-3}$  \\
                              & M & $0.95\pm0.07$ & $0.79\pm0.10$ & $0.16\pm0.06$ & $0.1\pm0.08$ & $<0.03$ & $<0.01$ & $<0.01$\\
\hline
Mg                            & R & $0.49\pm0.06$ & $0.45\pm0.06$ & $0.45\pm0.07$ & $0.30\pm0.08$ & $0.22\pm0.08$ & $0.41\pm0.09$ & $0.57\pm0.11$ \\
                              & M & $0.41\pm0.04$ & $0.43\pm0.06$ & $0.29\pm0.04$ & $0.26\pm0.06$ & $0.15\pm0.07$ & $<0.07$ & $<0.03$ \\
\hline
Si                            & R & $0.76\pm0.03$ & $0.78\pm0.03$ & $0.68\pm0.03$ & $0.54\pm0.04$ & $0.48\pm0.04$ & $0.39\pm0.04$ & $0.37\pm0.05$\\
                              & M & $0.71\pm0.02$ & $0.76\pm0.03$ & $0.61\pm0.02$ & $0.53\pm0.03$ & $0.46\pm0.04$ & $0.23\pm0.03$ & $<0.09$ \\
\hline
S                             & R &  $0.70\pm0.03$ & $0.68\pm0.03$ & $0.53\pm0.03$ & $0.40\pm0.04$ & $0.35\pm0.05$ & $0.20\pm0.05$ & $ <0.15$ \\
                              & M & $0.67\pm0.03$ & $0.65\pm0.04$ & $0.51\pm0.03$ & $0.38\pm0.05$ & $0.35\pm0.06$ & $0.12\pm0.07$ & $<0.10$ \\
 \hline
Ar                            & R & $0.66\pm0.06$ & $0.55\pm0.07$ & $0.35\pm0.08$ & $0.20\pm0.10$ & $0.32\pm0.12$ & $<0.21$ & $<0.04$\\
                              & M & $0.69\pm0.06$ & $0.50\pm0.09$ & $0.35\pm0.08$ & $0.19\pm0.12$& $0.24\pm0.14$ & $<0.18$ & $<0.07$\\
\hline
Ca                            & R & $0.74\pm0.10$ & $0.97\pm0.11$ & $0.98\pm0.13$ & $0.79\pm0.17$ & $0.78\pm0.18$ & $0.53\pm0.15$ & $<0.25$ \\
                              & M & $0.74\pm0.09$ & $0.98\pm0.11$ & $1.02\pm0.10$ & $0.77\pm0.15$ & $0.74\pm0.17$ & $0.66\pm0.17$ & $0.70\pm0.23$\\
\hline
Fe                            & R & $0.67\pm0.01$ & $0.63\pm0.01$ & $0.52\pm0.01$ & $0.44\pm0.01$ & $0.40\pm0.01$ & $0.36\pm0.01$ & $0.37\pm0.02$ \\
                              & M & $0.67\pm0.01$ & $0.62\pm0.01$ & $0.55\pm0.01$ & $0.43\pm0.01$ & $0.40\pm0.01$ & $0.31\pm0.01$ & $0.26\pm0.01$ \\
\hline
Ni                            & R & $1.70\pm0.13$ & $1.60\pm0.14$ & $0.73\pm0.16$ & $0.73\pm0.18$ & $0.76\pm0.20$ & $0.94\pm0.22$ & $1.9\pm0.3$ \\
                              & M & $1.65\pm0.10$ & $1.52\pm0.14$ & $0.99\pm0.10$ & $0.64\pm0.20$ & $0.69\pm0.17$ & un. & un.\\
\hline
$\chi^{2}$ / d.o.f.           & R & 726/525 & 764/525 & 814/525 & 721/525 & 653/525 & 712/499 & 1281/499\\
                              & M & 782/525 & 664/525 & 707/525 & 596/525 & 532/525 & 607/499 & 923/499\\ 
\hline

\end{tabular}
\noindent
\end{center}
\end{table*}

We compare the radial temperature and abundance profiles obtained using the combined background event lists 
of \citet{read2003} with those determined using the background model described in subsection \ref{backgmodels}.
We compare the values obtained using the {\it{wdem}} model, which was shown to describe the spectrum in the cool cores of clusters well (see Sect.~\ref{models}). We
derive a radial hydrogen column density profile, maximum temperature profile, $\alpha$ profile and abundance profiles of several elements: magnesium, silicon, sulfur,
argon, calcium, iron and nickel. 

Fitting the data using the background event files of \citet{read2003} we find an average hydrogen column density of $(2.49\pm0.03)\times10^{21}$ cm$^{-2}$, which is
significantly higher than the Galactic value of $1.80\times10^{21}$ cm$^{-2}$ determined from the 21 cm radio emission \citep{dickey1990}. This high absorption column density
was found before in ROSAT PSPC spectra by \citet{Irwin1995} and in a previous, shorter XMM-Newton observation by \citet{kaastra2003a}. The additional absorption might be caused
by Galactic dust clouds or molecular clouds. The measured radial absorption distribution has a slight peak between 2$\arcmin$ and 3$\arcmin$, but the errobars are large and it
is still consistent with a flat distribution (see Fig. \ref{NH}). Therefore, when fitting the data with our background model, we fix the absorption to $2.5\times10^{21}$ cm$^{-2}$. 

The left top panel in Fig. \ref{fig:MvR} shows that the radial
temperature profiles derived with the background event files of \citet{read2003} and with our background model are consistent with each other up to 6$\arcmin$ from the core of
the cluster. The best fit abundance values are, however, consistent only up to $4\arcmin$ from the core and in the outer parts, where the background starts to play a more
important role the differences are large. This is shown in Fig.~\ref{fig:MvR} and Table~\ref{tab:MvR}. We see that using the new background modeling the iron, silicon and
magnesium abundances between $4\arcmin-9\arcmin$ from the core are lower than those derived using the standard background event files. The large magnesium abundances at the
outer radii obtained using the standard background event files are due to the incorrect subtraction of the instrumental aluminum lines. Abundances of other elements are
consistent within the errorbars.

\end{document}